\newcommand{\nus}{{\nu_s}}

\documentclass[onecolumn]{aastex631}
\usepackage{lineno}
\usepackage{hyperref}

\def\lsol{\hbox{$\hbox{L}_\odot$}}

\newcommand{\sgra}{Sgr~A*}

\begin{document}
\title{Simultaneous {\it JWST}, {\it NuSTAR}, and {\it VLA} Monitoring of Sgr A*:\\
A Unified  Picture of the Variable IR, X-ray and Radio Emission}

 
\author[0000-0001-8403-8548]{F. Yusef-Zadeh} 
\affiliation{Dept Physics and Astronomy, CIERA, Northwestern University, 2145 Sheridan Road, Evanston , IL 60207, USA
(zadeh@northwestern.edu)}

\author[0000-0001-8403-8548]{M. Wardle} 
\affiliation{School of Mathematical and Physical Sciences, Astrophysics and Space Technologies Research Center, 
Macquarie University, Sydney NSW 2109, Australia, (mark.wardle@mq.edu.au)}

\author[0000-0001-8403-8548]{R. G. Arendt} 
\affiliation{Code 665, NASA/GSFC, 8800 Greenbelt Road, Greenbelt, MD 20771, USA}
\affiliation{UMBC/CRESST 2 (Richard.G.Arendt@nasa.gov)}

\author[0000-0001-8403-8548]{C.~O. Heinke} 
\affiliation{Department of Physics, University of Alberta, Edmonton, AB, T6G 2E1, Canada (heinke@ualberta.ca)}

\author[0000-0002-7570-5596]{C.~J. Chandler}
\affiliation{National Radio Astronomy Observatory, P.O. Box O,  Socorro, NM 87801, USA (cchandle@nrao.edu)}

\author[0000-0001-8403-8548]{H. Bushouse} 
\affiliation{Space Telescope Science Institute, 3700 San Martin Drive, Baltimore, MD 21218 (bushouse@stsci.edu)}

\author[0000-0002-7570-5596]{G.~A. Moellenbrock$^6$}

\author[0000-0003-3503-3446]{Joseph M. Michail}
\affiliation{Center for Astrophysics $|$ Harvard \& Smithsonian, 60 Garden Street, Cambridge, MA 02138, USA (joseph.michail@cfa.harvard.edu)}

\begin{abstract} 
Flux variability is a fundamental channel of information from Sgr A* because of its direct probe of processes occurring within an accretion disk under strong gravity.  We present simultaneous JWST,
NuSTAR, and VLA observations of Sgr A* on 2024 Apr 05.  We report the detection of a strong X-ray flare with a duration of about 40 minutes and a luminosity of $5.2\times10^{35}$ erg/s coincident with a
bright near-IR (NIR) flare, and a brightening in radio about an hour later.  We investigate the candidate physical mechanisms for the X-ray flare emission and conclude that this can best be
explained by inverse Compton scattering of NIR flare radiation. We propose a dynamic scenario analogous to a coronal mass ejection in which a magnetic flux rope is ejected from Sgr A*'s inner
accretion flow with a current sheet extending down from the rope to the bulk of the accretion flow. The accelerated electrons are continually ejected from the reconnection X-point with a bulk flow
at the Alfven speed of 0.7c. IR radiation from the approaching energetic electrons is enhanced by beaming and up-scattered by thermal electrons in the accretion flow to produce the strong X-ray
flare.  Meanwhile, the relativistic electrons moving in the opposite direction away from the disk experience weaker magnetic fields so radiate at longer wavelengths. They feed into the magnetic
flux tube explaining the detected delayed radio emission. This physical picture attempts to unify the origin of the variable emission from Sgr A* at IR, X-ray and radio/submm wavelengths.
\end{abstract}

\section{Introduction} 
\label{sec:intro}

The 4$\times10^6$ solar mass black hole at the center of the Galaxy, \sgra, and its accretion flow provide a unique laboratory for studying exceptional physical processes 
\citep{genzel10}. Since \sgra\ was first discovered by \cite{balick74}, its emission has been studied extensively over a wide range of the electromagnetic spectrum, from radio to 
X-rays and $\gamma$-rays, with the exception of the optical and ultraviolet bands due to high extinction 
\citep{falcke98,melia01,baganoff03,cafardo21,zhao03,hess16,magic20,adams21,cafardo21}.  \sgra is over 100 times closer than the next nearest supermassive black hole, offering an 
unparalleled opportunity to observe its behavior and structural details.  The bolometric luminosity of \sgra\ is $\sim$100 \lsol\, about 10$^{-9}$ times its Eddington luminosity 
\citep{gillessen09,genzel10}. \sgra\ is now identified as one of the weakest accreting black holes accessible to detailed observations, thus giving us great opportunity to understand the 
nature of black hole emissions in the limit of weak accretion.  Sgr A*'s radiation is unprocessed and directly arises from the accretion flow and/or 
small-scale jet, thus models of its emission can be examined in detail. The accretion flow of Sgr A* is thought to be fueled by winds from young mass-losing stars orbiting within 0.1 
pc. The exact nature of the accretion flow and its possible jet emission, however, is still under debate \citep{yuan03,quataert04,cuadra06,yuan14,ressler20a,ressler20b,wielgus22}.

Its extreme faintness compared to its Eddington luminosity suggests that \sgra\ accretes matter through a geometrically-thick radiatively inefficient accretion flow (RIAF) rather 
than a thin disk \citep{narayan94,yuan03,yuan14}. Its radio spectrum resembles the jet emission seen in active galactic nuclei, raising the possibility that its radiation originates 
from a jet rather than direct accretion \citep[e.g.][]{falcke00}.  Recent GRMHD simulations produce jets, outflows, and variability, \citep[e.g.][]{ressler23}. These simulations 
focus on global structure of the accretion flow and reveal 
a magnetically arrested inner disk (MAD) (strong, organized fields that can halt accretion) \citep{ressler20a}.


The emission from the accretion flow is variable. There have been numerous multi-wavelength studies of flaring activity of Sgr A* over the last 20 years 
\citep{baganoff03,genzel03,porquet03,goldwurm03,ghez04,zadeh06a,gillessen09,dodds-eden09,dodds-eden11,nowak12,degenaar12,neilsen15,bower15a,ponti15,mossoux16,karssen17,witzel18,do19,eckart19,michail21,wielgus22,weldon23,michail24,fellenberg25,zadeh25}.  
For the first time, the GRAVITY instrument \citep{gravity18} tracked relativistic, periodic motion of the centroid of the NIR flare emission. The emission arises 
from within several gravitational radii, $r_g = GM/c^2$, suggesting orbital motion around Sgr A*. This striking result indicated that flaring activity of Sgr A* is 
a direct probe of the physical processes under strong gravity that likely stem from gas dynamical flow and radiation in the accretion disk. A number of simulations 
suggest that the particle acceleration is due to magnetic field reconnection, MHD turbulence, shocks, or jets 
\citep{markoff01,yuan03,igumenshchev08,yuan09,dodds-eden10,dibi14,dexter20,ripperda20,ressler20a,ball21,porth21,cemeljic22,aimar23,lin24,jiang24}.  
Tidal disruption of 
asteroids has also been suggested as the origin of flares \citep{cadez08,kostic09,zubovas12}.

JWST's ability to conduct long duration, high-resolution monitoring at two NIR wavelengths has opened up a great opportunity to compare NIR 
variability to the emission at other wavelengths.  JWST, NuSTAR and VLA observed Sgr A* simultaneously at two epochs, on April 5 and April 7, 2024. 
There was a powerful X-ray flare coincident with a NIR flare on April 5. The X-ray observations did not overlap with the time when a NIR flare was 
detected on April 7, 2024. We also carried out contemporaneous VLA observations at Ka band (34 GHz) and detected a broad radio peak in the light 
curve of Sgr A* that was delayed with respect to the NIR flares on April 5.

The X-ray flare on April 5, 2024 is one of the strongest detected, with a (2-10) keV luminosity of L$_{\rm {x}}\sim 5.2\times10^{35}$ erg\, s$^{-1}$, comparable to the bolometric 
luminosity of Sgr A* and about two 
orders of magnitudes brighter than the quiescent X-ray luminosity, which is thought to be produced on scales $\sim 10^5 r_g $ 
by thermal bremsstrahlung from captured stellar wind material \citep{cuadra06,cuadra08,ressler20a}.

In this paper, we first summarize past NIR, X-ray and radio/submm measurements of Sgr A* in $\S2$, followed by an overview of some recent simulations of the accretion flow of Sgr A*. 
After describing details of our contemporaneous observations using JWST, NuSTAR and VLA in $\S3$, we present results of our measurements at NIR, X-ray and radio in $\S4$,
including  accurate characterization of the spectral evolution of NIR, and radio flare emission and time evolution of X-ray flare emission.  
A model of radio emission is described in $\S4.3$ and  synchrotron, ICS and SSC models of X-ray emission are discussed in $\S5$. 
We conclude that the only means of producing the intense X-ray emission is through inverse Compton scattering (ICS) of the IR flare emission, but 
only if the IR emission is upscattered by thermal  electrons in the disk.  In $\S6$,  we propose a scenario in which 
relativistic electrons  are accelerated in the reconnecting region  and are pushed downwards 
towards the disk with bulk  velocity close to the  Alfven speed which is a natural consequence of reconnection. 
Transient relativistic electrons produce  strong IR radiation, as viewed by the disk, before  IR emission  being 
upscattered by  thermal electrons in the disk.  In $\S7$, we provide a summary of our model of the origin of variable 
IR, X-ray and radio/submm emission. 


\vfill\eject
\section{Past Studies}

\subsection{JWST Observations of Sgr A*}

A number of ground- and space-based studies have examined the total IR intensity and polarization variability of Sgr A*, \citep[e.g.][]{
genzel03,ghez04,meyer06,trippe07,zamaninasab08,marrone08,zadeh08,nishiyama09,gillessen09,dodds-eden09,hora14,mossoux16,witzel18,do19,eckart19,wielgus22,michail24,fellenberg25}.  
At NIR wavelengths, recent long and multi-epoch JWST measurements of Sgr A* at two different wavelengths characterized the spectral evolution of flares over 7 
epochs in 2023 and 2024 \citep{zadeh25}. Multi-epoch NIR observations for a total of 48h traced the light curves of Sgr A* (over a range of days to months) and 
indicated that Sgr A* has three states, hereafter ``flickering'', ``flaring'' and ``quiescent'' states: (i) nonstop continuous flickering on few minute timescales; 
(ii)  powerful flares on timescales of hours, a few times a day;  and (iii) a steady low-level pedestal, or ``quiescent'' component. 
Unlike the first two states that 
are transient and vary on short time scales, the quiescent state shows day-to-day, and month-to-month, variability in color or spectral index, as well as brightness 
variations. The exact flux of the pedestal is subject to details of the background subtraction \citep[see details in ][]{zadeh25}. The pedestal emission arising from 
Sgr A* is of the order of $\sim0.1-1$ mJy at 2.1 $\mu$m, corresponding to $\sim (0.1-10)\times10^{34}$ erg s$^{-1}$. 
We consider this to be   the 
quiescent state of Sgr A* at IR wavelengths, placing constraints on the SED of Sgr A*.  

Evidence for the reality, and association with Sgr A*, of the 
background subtracted low-level flux is that the emission varies slowly from epoch to epoch.  Additionally, the spectral index of the flickering would vary in 
unrealistic ways if there were no pedestal so that the flux went to zero at the faintest excursions of the flickering.  The submillimeter bump in the SED of Sgr A* 
is due to thermal synchrotron emission from the disk, and the IR excess traced by the pedestal/quiescent flux is likely to be due to a nonthermal high-energy tail 
in the electron distribution function.

The comparison of the 2.1 $\mu$m light curve with the NIR spectral index based on JWST data indicated steepening of the spectral index  as 
the NIR flares become brighter. The loop pattern of the light curves showed a  time delay and steepening between 4.8 and 2.1$\mu$m emission, 
due to synchrotron cooling. This allowed estimates of flare  magnetic fields. For example,   
the magnetic field was estimated  to be 88.5 G for a flare that peaked at 5.25h on April 5, 2024 \citep{zadeh25}.  
The steepening occurs for the relatively faint flares present on most days. However, 
for the brightest flares (those that get above the $``break''$  in slope" in the  2.1 vs 4.8 $\mu$m  flux plot) \citep{zadeh25}, 
the spectral index gets flatter with brightness. 
This happens for the bright flares on April 5 and 7, 2024 \citep{zadeh25}.


Plots of the 2.1 $\mu$m flux vs 4.8 $\mu$m flux density of Sgr A* showed a break in slope between dim flickering emission and the brighter NIR flares, indicating 
two distinct populations of energetic electrons \citep{zadeh25}. Properties of  flickering and flaring states of Sgr A*  are consistent with earlier NIR studies,
\citep[e.g.][]{dodds-eden11}. 
The minute time-scale flickering component is thought to arise from synchrotron emission of electron energies with  power-law tail extending from the thermal 
electron population in the inner accretion flow, perhaps moderated by turbulent fluctuations in density and magnetic fields, e.g., \citep{grigorian24}.  The flaring 
component in the NIR is polarized \citep{eckart06}, so is certainly synchrotron emission from a transient population of electrons, likely accelerated by magnetic 
reconnection during the sporadic magnetic restructuring events seen in the simulations of magnetically-dominated accretion disks \citep{ripperda20,ressler20a}.

\subsection{X-ray Variability}

At X-ray energies, numerous studies have focused on the flare and quiescent emission and the relationship between NIR and X-ray hourly timescale flare emission 
\citep{liu02,baganoff03,genzel03,ghez04,goldwurm03,porquet03,eckart04,eckart06,belanger05,porquet08,dodds-eden09,trap11,nowak12,degenaar12,barriere14,yuan14,neilsen15,mossoux16,ponti17,boyce18,boyce19,haggard19,boyce21,gravity21}.  
The quiescent component of the X-ray emission from Sgr A* is partially resolved on the scale of the Bondi radius $\sim1''\, (2\times 10^5 r_g$)  with a luminosity 
of $\sim3.6\times10^{33}$ erg s$^{-1}$ \citep{baganoff03}. Thermal bremsstrahlung emission from hot plasma with a temperature of $\sim7\times10^7\,$K and electron density 
$n_e\sim100$ cm$^{-3}$ can explain the quiescent component, either from infalling stellar wind material captured 
by Sgr A* \citep[see][]{quataert02} or alternatively from coronal gas associated with a population of spun-up stellar sources near Sgr A*, e.g., \cite{sazonov12}.

Detectable X-ray flares have a lower duty cycle than detectable NIR flares, at a rate of $\sim1-1.3$ flares day$^{-1}$ \citep{neilsen15}. X-ray flare emission is 
considered to be due to either synchrotron emission, inverse Compton scattering (ICS), or synchrotron-self Compton (SSC) emission, e.g., \cite{ponti17,haggard19}. 
The peak X-ray flare emission can be 1-100 times more luminous than the quiescent X-ray luminosity. X-ray flares always show NIR counterparts, suggesting that they 
have a similar origin. It is, however, not clear why certain NIR flares have X-ray counterparts and some do not \citep[e.g.][]{horn07}.  The brightest X-ray flare 
that has been detected between 2-10 keV reached a peak of $L_{\rm {X}} =1.2\times10^{36}$ erg s$^{-1}$ \citep{haggard19}.  Three of the other brightest X-ray flares 
have a range of luminosities between $(1-5)\times10^{35}$ erg s$^{-1}$ \citep{porquet08,nowak12,zhang17}. 
In spite of a large body of literature on this subject, 
there is no consensus on the origin and radiation mechanism responsible for the production of X-ray flare emission.


\subsection{Radio/Submm  Variability}

Sgr~A* was first discovered in 1974 and its radio flux variability was identified within a few years \citep{balick74,brown82}. The size of the radio source's photosphere decreases with increasing frequency 
\citep{bower14}. The majority of the luminosity of Sgr A*, $5\times10^{35}$ erg s$^{-1}$, is detected at submm and radio wavelengths \citep{bower19}. The flux density rises at higher frequencies and peaks at the 
``submillimeter bump" around 350 GHz before falling steeply.

The steady radio/submm component is primarily due to optically thick synchrotron radiation produced by relativistic thermal electrons having a temperature of a few $10^{10}$K and $n_e\sim10^{6-7}$ cm$^{-3}$ embedded 
in a magnetic field of $\sim10-50$ G \citep{loeb07,genzel10}. The radio emission is emitted at $10^{3-4}r_g$, while emission at frequencies above the submm bump is dominated by optically thin emission closest to the 
black hole at $\sim10-100\,r_g$ \citep{falcke98,fazio18}. The inner $\sim 1-2''$ of Sgr A* also shows a steady source of diffuse millimeter emission surrounding Sgr A*. This has been argued to be produced by 
synchrotron emission from relativistic electrons in equipartition with a  $\sim 1.5$ mG magnetic field; the origin of this feature is not clear but its coexistence with hot gas suggests an outflow from the ionized 
winds of the S stars orbiting Sgr A*, or from Sgr A* \citep{quataert05,zadeh17}.

Sgr A* is also variable at radio/submm wavelengths, though the flux is dominated by the quiescent component 
\citep{zhao03,mauerhan05,marrone08,doeleman08,eckart08,zadeh09,li09,devaky10,sabha10,fish11,michail21,michail24}.  At radio and submm wavelengths, the 1-2h flare duration strongly suggests that the decline is due to 
adiabatic cooling of an expanding synchrotron hotspot.  The frequency-dependent radio and submm time delay \citep{zadeh06a} is consistent with an initially optically thick hotspot in the accretion flow.  As the 
hotspot expands, the intensity increases, peaks, and declines at successively lower frequency as it transitions from optically thick to thin \citep{vanderlaan66,zadeh06a,brinkerink16,michail21}. Arguments have been 
made that radio/submm variable emission is time-delayed with respect to NIR flare emission, implying that perhaps one population of relativistic electrons can explain the variable emission at NIR as well as at 
radio/submm. Submm/radio variable emission has also been interpreted as arising from a collimated outflow produced from shocks or magnetic reconnection within a jet 
\citep{markoff01,yuan03,zadeh06b,marrone08,zadeh08,dodds-eden09,rauch16,lin24}.

Observations and simulations of the variability on minute-timescale have also been reported at radio wavelengths \citep{zadeh11,murchikova22}.  Flickering at radio wavelengths is 
thought to arise from the expansion of optically thick synchrotron emitting plasma regions, suggesting the presence of explosive, energetic, relativistic expansion events 
\citep{zadeh11}.  These events could feed the base of an outflow or jet in Sgr A*, as many recent GRMHD simulations indicate. VLBI observations of Sgr A* suggest that the morphology 
of Sgr A* at 43 GHz is changed slightly $\sim$4.5h after a NIR flare \citep{rauch16}.

\subsection{Simulations and Calculations}

A large number of theoretical and numerical studies have recently examined the nature of Sgr A*'s flaring activity close to the event horizon. Many studies, 
including detailed multi-stage simulations, show that colliding winds originating from orbiting massive, young stars feed Sgr A*, and conclude that the 
amplification of the weak stellar wind fields is sufficient that a MAD disk is the most plausible model for the inner accretion flow 
\citep{liu02,igumenshchev08,dodds-eden10,dibi14,ball18a,ball18b,dexter20,ripperda20,ressler20a,petersen20,ball21,porth21,cemeljic22,nathanail22,aimar23,lin24}. 
In 
the MAD model, the poloidal component of the magnetic field piles up by the accreting gas close to the inner disk where the magnetic pressure and gas pressure 
balance each other.  MHD simulations support a picture in which magnetic flux eruptions of hot spots in the disk drive flares. This is supported by GRAVITY 
polarization measurements, which indicate that the field in NIR flares is predominantly poloidal \citep{gravity18}. Furthermore, these simulations also find current 
sheets in the disk with frequent eruptions of excess magnetic flux from the inner accretion disk supporting the idea that magnetic reconnection is able to 
accelerate particles to high energies and cause the variability and flaring of Sgr A* in the NIR 
\citep{gutierrez20,chatter21,nathanail22,aimar23,dimitro24,scepi24,antono25}.  
The strong hour-long flaring component in the NIR is consistent with non-thermal 
electrons being accelerated by occasional reconnection events within the bulk of the accretion flow \citep[e.g.][]{scepi24,grigorian24} or, alternatively, from 
reconnection-driven ejection of plasmoids out of the disk plane \citep[e.g.][]{aimar23,jiang24,jiang25}. These simulations are consistent with a picture of expanding hot spot 
eruptions that had been inferred from frequency-dependent time delays of the variable radio emission \citep{zadeh06a,brinkerink16}; see more details in $\S4.2$ and $\S4.3$.

\begin{figure} 
   \centering
   \includegraphics[width=2.6in]{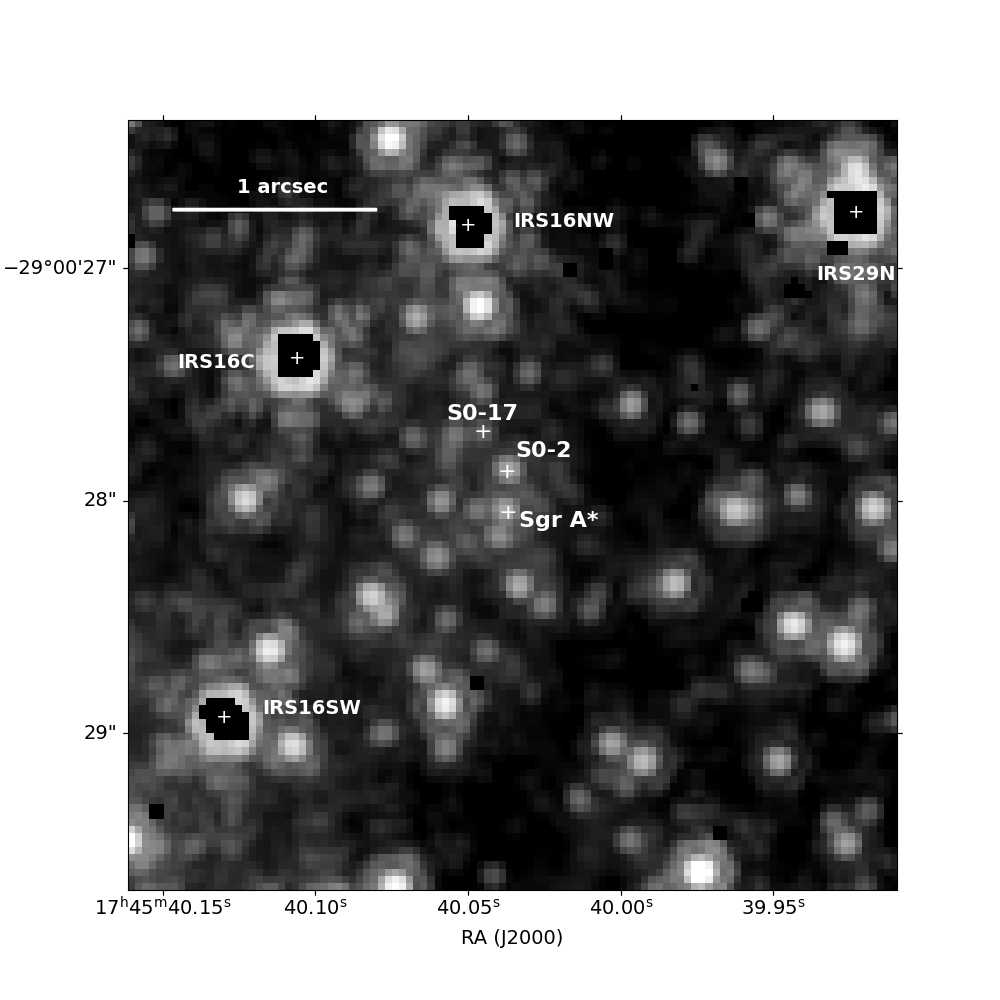}
   \includegraphics[width=2.6in]{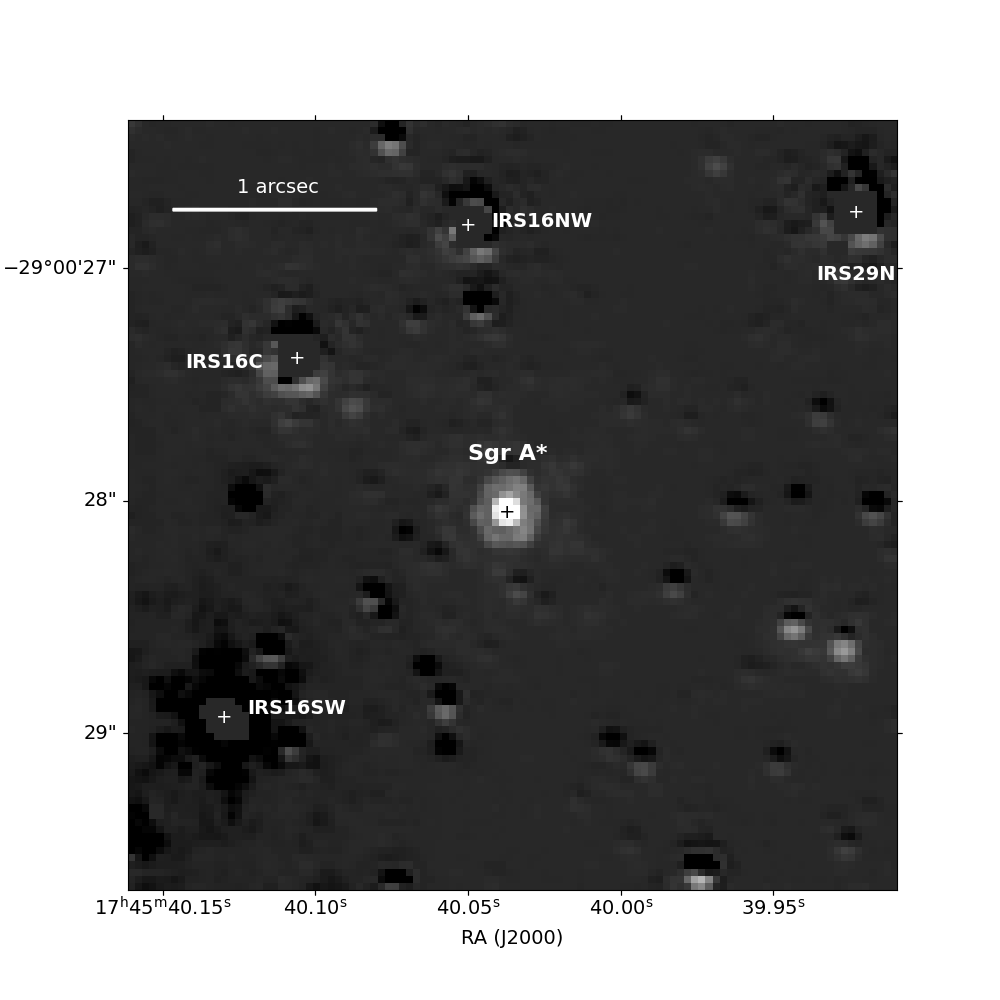}\\ 
\vspace{-0.2in}   
\includegraphics[width=5in]{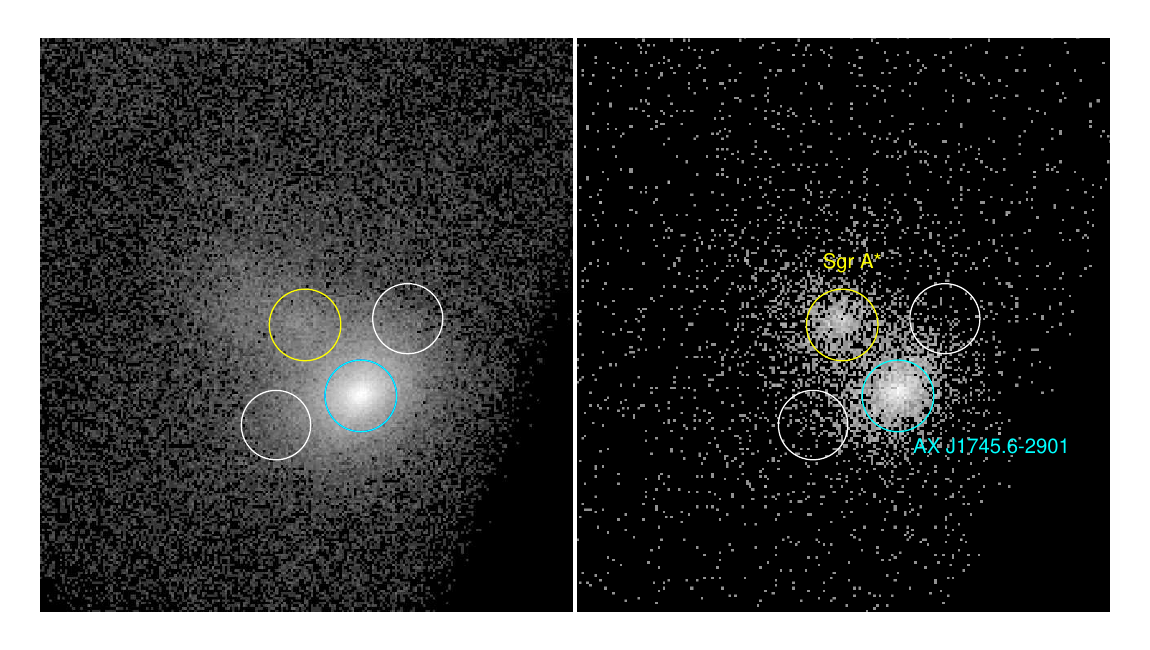}\\ 
\vspace{-0.1in}   
   \includegraphics[width=4.35in]{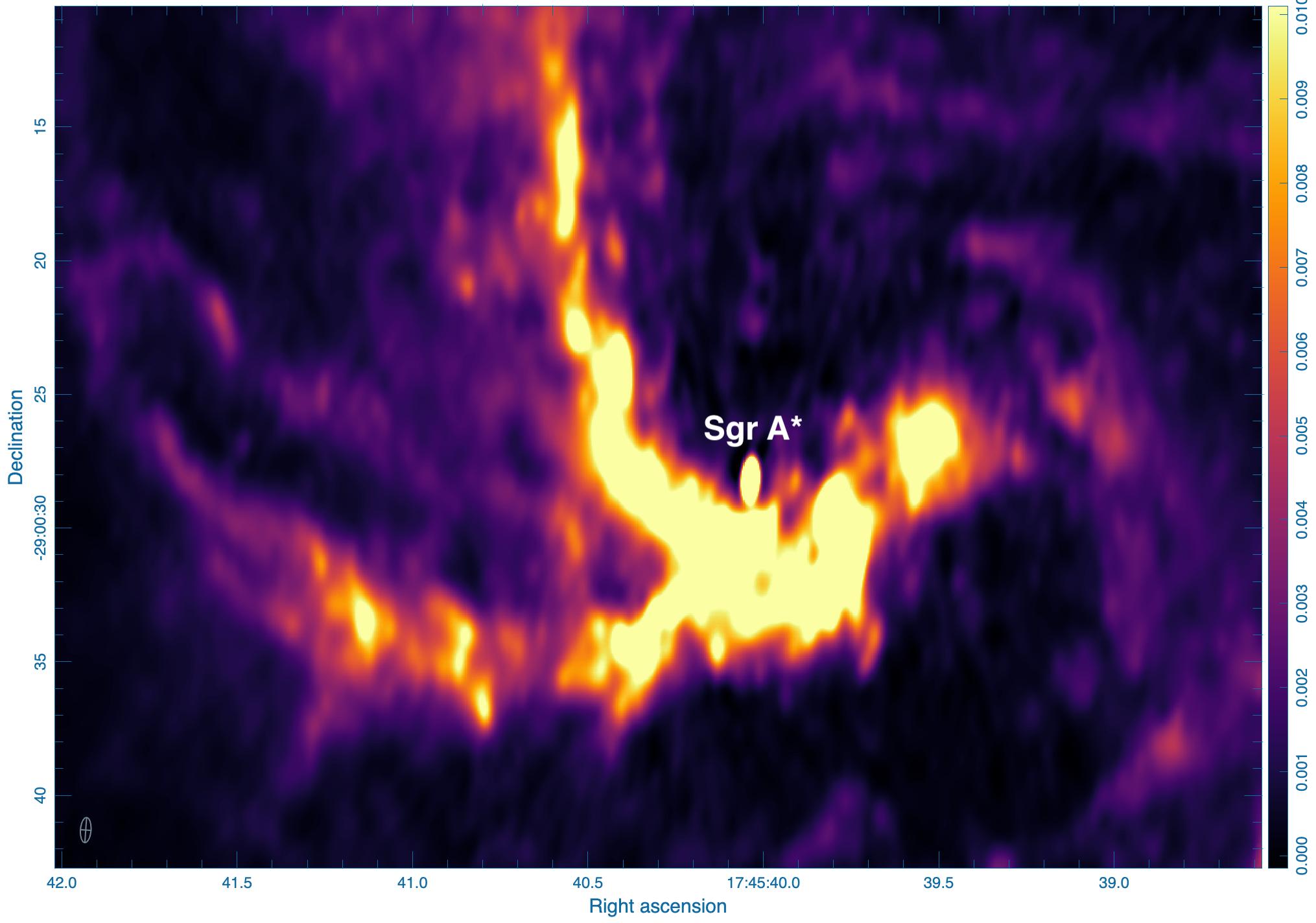}\\ 
 \caption{
{\it (a)}
The normal stellar field at 2.1 $\mu$m of NIRCam of JWST surrounding Sgr A* (top left), whereas a version 
that has had a baseline image (when Sgr A* was not flaring) subtracted from it (top right).  
This allows Sgr A* to be seen without the overlapping nearby stars. Note that the cores of bright stars in the 
field, such as the IRS16 members and IRS29N, are saturated and hence appear black.
{\it (b)}
An X-ray image showing $34.7''$ extraction circles around Sgr A* (yellow), the nearby bright LMXB AX J1745.6-2901 (cyan), and 
the background regions used for Sgr A* (white, chosen to be equidistant from the LMXB as Sgr A*. 
The images show when Sgr A* is not flaring (middle left) and when flaring (middle right, shorter 1000 s exposure time).
{\it (c)}
A VLA image of the inner $\sim30\times40''$ of Sgr A* showing the mini-spiral ionized gas at 34 GHz (bottom), 
taken simultaneously on April 5, 2024.  The bright compact source is Sgr A*.  At this frequency 
the quiescent flux 
dominates over the variable emission.
}
\end{figure}

\section{Observations}

Here we describe details of JWST,  NuSTAR, and VLA observations that were taken simultaneously on 
April 5, 2024 when  a powerful NIR and X-ray flare was detected.

\subsection{JWST}

Recent JWST observations of \sgra\ have been reported in detail \citep{zadeh25}.  Briefly, NIR photometric monitoring data of Sgr A* were obtained with the NIRCam 
instrument of JWST at  7 epochs from April 2023 through April 2024.  Simultaneous imaging was obtained in the NIRCam short- and long-wavelength channels, using the 
medium band filters F210M and F480M. These filters have central wavelengths of 2.096 and 4.814 $\mu$m, with bandwidths of 0.205 and 0.299 $\mu$m, respectively.  
Observations on April 5, 2024 used the NIRCam Time Series Imaging method, where the source was left at a fixed location over the hours-long duration of the 
observation. Thousands of short exposures were obtained over $\sim$8 hours and photometry of Sgr A* in each exposure was used to produce continuous light curves.

Figure 1a shows  JWST NIRCam 2.1$\mu$m images of the Galactic center region (see details of extinction correction in \citep{zadeh25}) 
when Sgr A* exhibited a bright flare at 10.5 hrs UT on 5 Apr, 2024 and 
a difference image between the flaring and non-flaring states. The locations of well-known members 
of the stellar cluster near Sgr A*, such as S0-2, are shown. 
Extinction corrections of
$A_{K_s}= 2.46\pm0.03$ (for F210M) at 2.1$\mu$m and  $A_M = 1.0\pm0.3$ (for F480M) have been applied, as stated in 
\citet{schodel11}. 

\subsection{NuSTAR   X-ray data reduction} 

The NuSTAR observations were taken simultaneously with JWST during two epochs, 
April 5 (3:43 to 15:08 UT) and April 7 (3:51 to 15:21 UT), 2024.  We identified no 
variability in the X-ray emission from Sgr A* on April 7, 2024 so we focus on the April 5, 2024 data. The X-ray data were barycentered with the position of Sgr A* 
using the FTOOL $``axbary"$, and products were extracted using the standard ``nuproducts" FTOOL. We chose a circular region of $34.7''$ in size around Sgr A* to extract 
products, chosen to include Sgr A* light and exclude most light from the active nearby LMXB AX J1745.6-2901,  see e.g. \cite{degenaar12,reynolds23}, which is 
brighter than Sgr A* and only 1.5$'$ away (see Fig.\ 1b). We selected two background regions spaced around AX J1745.6-2901 at the same distance as Sgr A*. Stray 
light was visible at the northeast corner of the field of view, but did not affect the regions surrounding Sgr A*. We focus on data from module B, which has less 
background contamination from stray light.

We binned the lightcurves in intervals of  80.46 s, and computed Poisson errors on the countrates using \cite{Gehrels86}. Using the spectral fit below, we convert 1 NuSTAR 
(module B) ct s$^{-1}$ (in 3-79 keV) to an unabsorbed flux of $3.8\times10^{-11}$ erg/cm$^2$/s (3-10 keV), and then convert that to a mean flux density of 2.26 $\mu$Jy 
over the equivalent frequency range. Our April 5 lightcurves show a clear flare in both NuSTAR detectors (see $\S4.1$). To verify that this flare is from Sgr A* and 
not from AX J1745.6-2901, we also extracted an 
April 5 lightcurve from AX J1745.6-2901, which shows no flare, but one clear eclipse as previously seen from AX 
J1745.6-2901 \citep{maeda96}, confirming our interpretation.

We also fit the X-ray spectrum from the April 5 observation as a whole with an absorbed power-law, finding $N_H=1.1\pm0.3\times10^{23}$ cm$^{-2}$, photon index 2.56$\pm0.17$, $F_X=1.8\pm0.3\times10^{-11}$ 
erg  s$^{-1}\, {\rm cm^{-2}}$. An Fe K line is clearly visible, but it is unclear whether this originates from Sgr A* or from the Sgr A East SNR, which cannot be disentangled from Sgr A* at the spatial resolution of NuSTAR 
(43$''$). However, we can separate the X-ray flare emission from Sgr A* by using the non-flare emission as background.

We simultaneously fit the X-ray spectrum of 1000 seconds of flare, and of the remaining non-flare spectrum, which we take to also be present in the flare spectrum. Unless otherwise stated, quoted errors in this 
section are 1$\sigma$. After iteration, we fit an absorbed power-law ($N_H=2.5\pm2.5\times10^{22}$ cm$^{-2}$, photon index 2.0$\pm0.2$, $F_X$(2-10 keV)$=7\times10^{-12}$ ergs cm$^{-2}$ s$^{-1}$), plus a gaussian (at 
6.5$\pm0.1$ keV, equivalent width 1.0$\pm$0.4 keV at 90\% confidence) to the non-flare spectrum. This spectrum includes flux from Sgr A*, as well as flux from multiple nearby sources, including the Sgr A East 
supernova remnant, \citep[see e.g.][]{baganoff03}. Then we add an absorbed power-law to this model to fit the flare spectrum. We find an additional $N_H$ of $1.5^{+1.1}_{-0.8}\times10^{23}$ cm$^{-2}$ for the flare, 
with power-law index of 2.1$\pm0.3$ and 2-10 keV flux of $7^{+3}_{-2}\times10^{-11}$ ergs cm$^{-2}$ s$^{-1}$. We also fit the first and second half of this flare separately, and find values of the photon index and 
normalization that are consistent within the errors.

\subsection{VLA and radio data reduction}

The 1-cm data were obtained using the Karl G. Jansky Very Large Array (VLA) on 2024 
April 5 under project code VLA/24A-106, when the VLA was in the C-configuration 
(maximum baseline 3.4 km). The correlator was configured to observe from 28.975 GHz to 37.023 GHz using 64 spectral windows, each covering 128 MHz with 64 2-MHz 
channels. The bandpass calibrator was J1733$-$1304, and the complex gain calibrator was J1744$-$3116. The absolute flux density scale was obtained from observations 
of 3C286 (Perley \& Butler 2017), which was observed at a considerably higher elevation than Sgr A*. The overall uncertainty in the absolute flux density scale is 
therefore somewhat larger than typical, and is estimated to be 20\%. When imaging the data over all spectral windows, the flux densities of Sgr A* are 
1.34$^{+0.21}_{-0.15}$ Jy centered at 33 GHz.

The data were calibrated using the standard VLA calibration pipeline, version 2024.1.0.8. 
 Additional flagging of bad data was carried out by hand. In order to obtain the light curve 
for Sgr A*, one round of 
phase-only self-calibration using a point-source model at the phase center was performed, after which the data were averaged per-spectral window. The data were then imported into AIPS, where the task DFTPL was used 
to plot the real part of the visibilities after time-averaging, to form the light curve.

The time variability of Sgr A$^*$ causes severe problems for interferometric imaging of the Galactic center, since Earth rotation synthesis assumes the sky emission 
remains constant while the rotation of the Earth is used to fill in the {\it uv}-plane. 
A time- and frequency-dependent model for Sgr A$^*$ therefore needed to be removed 
from the visibility data before attempting to image the emission from the rest of the field. Details of how this was achieved will be given elsewhere, 
but in summary, a constant point-source model for Sgr A$^*$ was used to derive a time-dependent bandpass correction, and applied to the data. 
This absorbs the variability 
of the Sgr A$^*$ point source, and applying this correction renders Sgr A$^*$ constant, permitting the model for Sgr A$^*$ to be subtracted from the visibilities. 
The bandpass correction is then reversed, leaving un-self-calibrated visibilities containing just emission from the mini-spiral. The time-dependent phase 
corrections derived from the bandpass correction were then applied to the mini-spiral visibilities. Antenna-based amplitude corrections were derived by normalizing 
antenna-based self-calibration gain solutions from Sgr A$^*$ per integration, to take out the variability of Sgr A$^*$, and also applied to the mini-spiral 
visibilities. After imaging, a (constant) point source was added back into the image to provide a reference for the position of Sgr A$^*$, as shown in Figure 1c.


\begin{figure}[htbp] 
   \centering
   \includegraphics[width=3.5in]{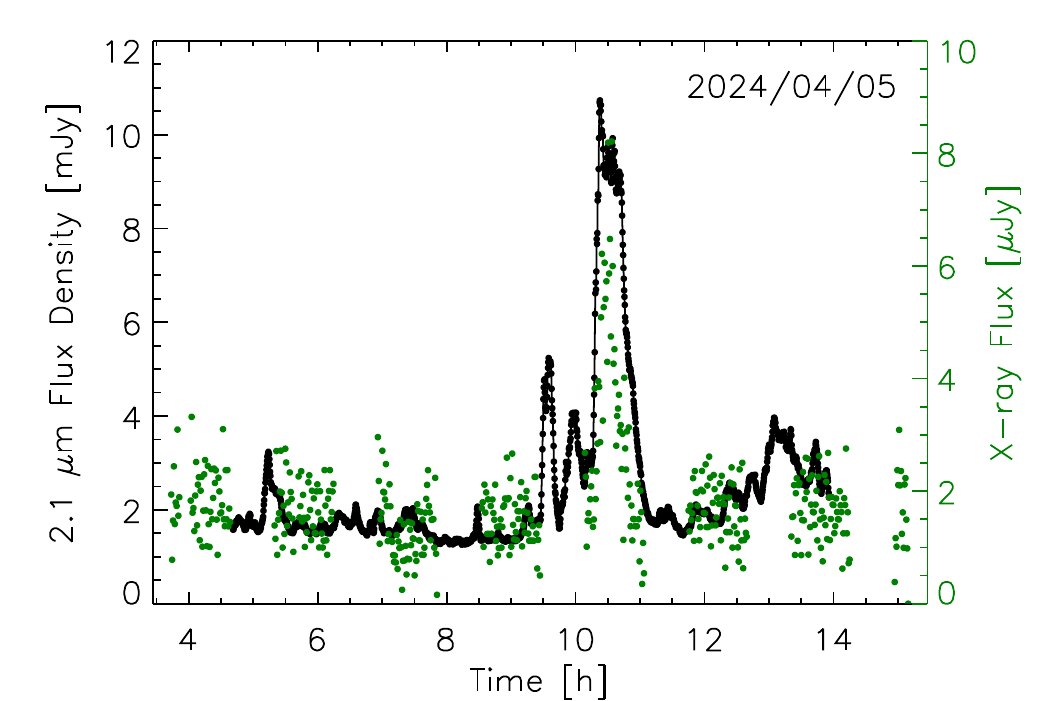}
   \includegraphics[width=3.5in]{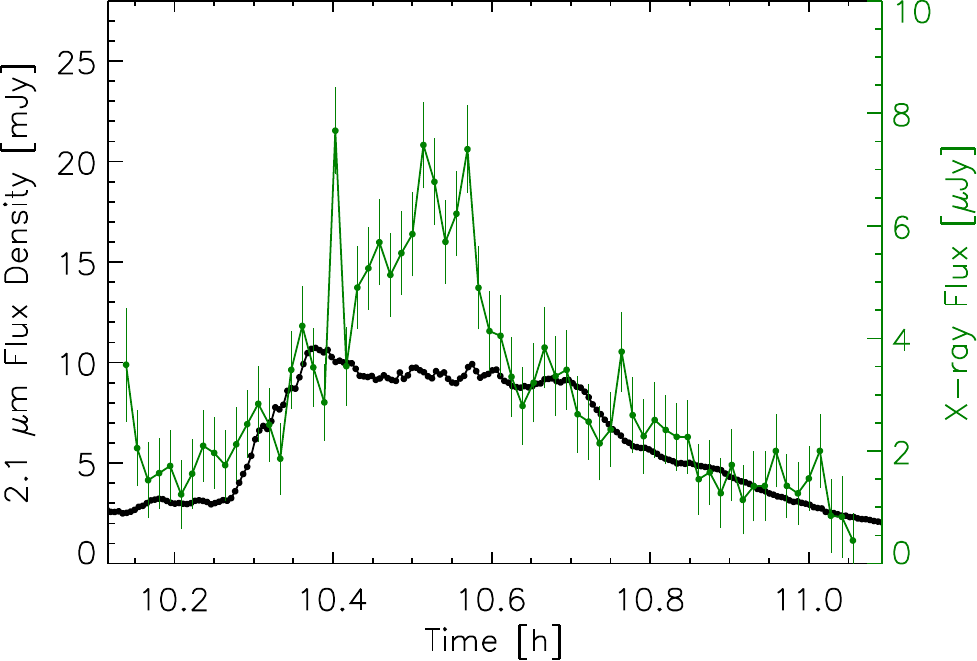} 
   \includegraphics[width=3.5in]{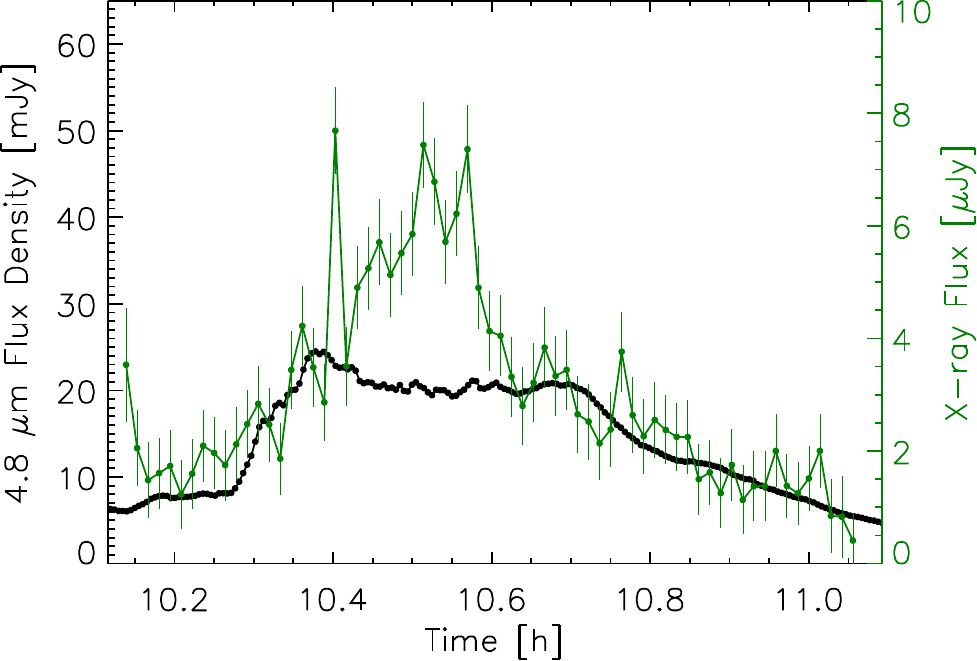}
 \caption{
{\it (a) Top Left}
The 2.1 $\mu$m (black) and 3-10 keV NuSTAR (green) light curves of   Sgr A* 
show a strong X-ray flare coincident with a NIR flare.  
{\it (b) Top Right}
A close-up view of  the NIR light curve of the flare at 2.1 $\mu$m is shown against the  X-ray light curve but normalized such that
both light curves rise together.   
The X-ray flare is detected within the envelope of the NIR flare. 
{\it (c) Bottom}
Similar to {\it (b)} except at 4.8$\mu$m.
}
\end{figure}

\section{Results}

\subsection{NIR vs  X-ray light curves}

Figure 2a shows a JWST light curve of Sgr A* at 2.1 $\mu$m (black)  superimposed with the 3-10 keV X-ray counterpart from NuSTAR (green). The JWST 2.1$\mu$m light 
curve shows continual variability on minute-to-hourly time scales, providing information on horizon-size scales.  A closer views of the light curve at 2.1 and 4.8 $\mu$m are 
shown in Figures\ 2b,c  where  the scales are normalized differently than Fig. 2a. 
We note the X-ray and IR flare emission have similar  duration;
both are correlated with each other as they  rise and become brighter until the  IR emission plateaus in brightness between 10.4-10.6h  UT 
whereas  the X-ray continues to rise to a sharper peak. The correlation is also noted when both IR and X-rays fall  together. 
It is not clear why IR flux saturates as this behavior is also seen on our April 7, 2024 epoch  observations \citep{zadeh25}. 
Cross correlation of the light curves shows no time delay between the X-ray and NIR flare emission.

\begin{figure}[htp] 
   \centering
      \includegraphics[width=3in]{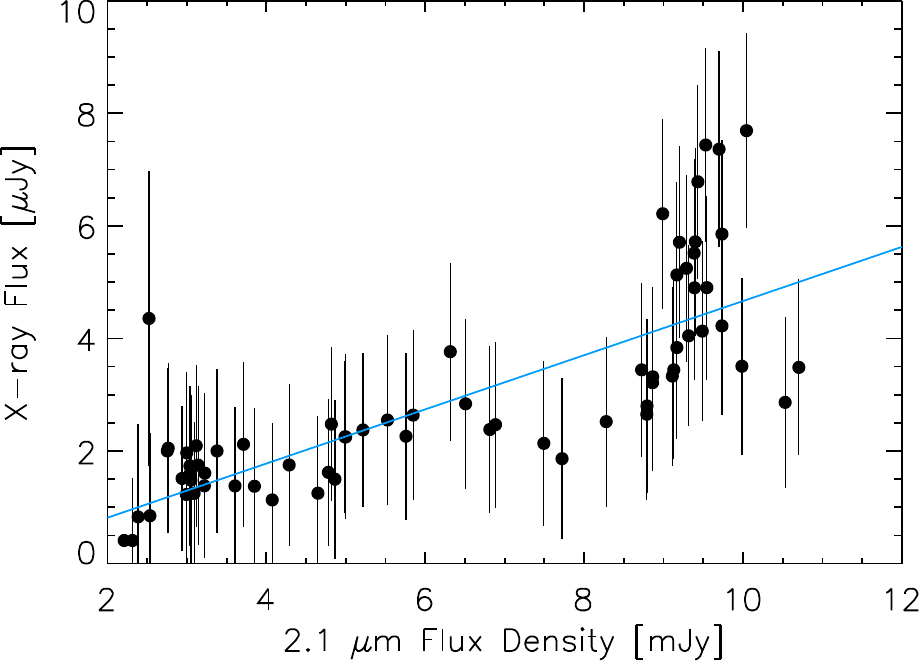} 
\vspace{0.2in}
   \includegraphics[width=3in]{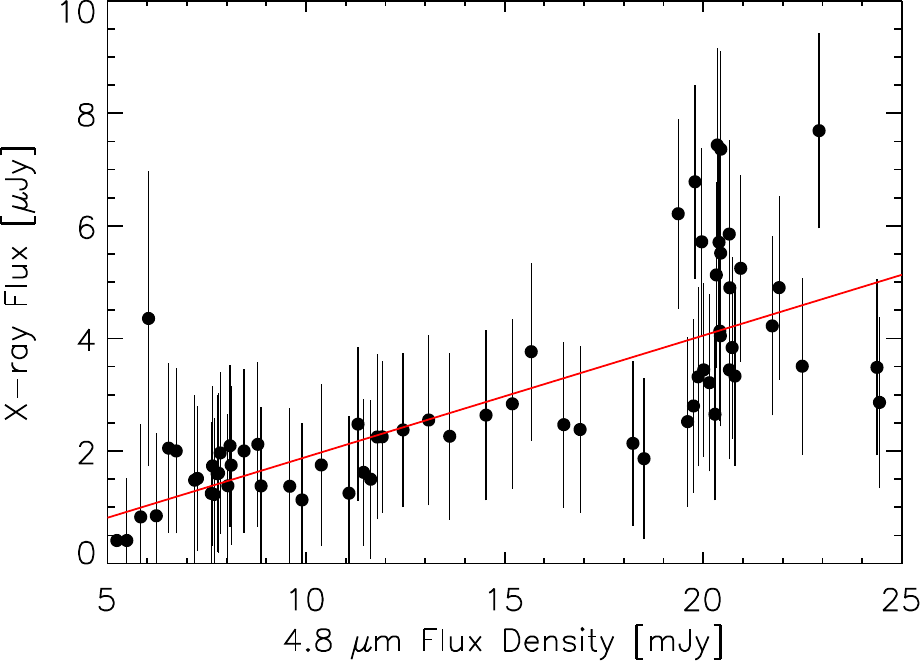} 
\vspace{0.2in}
      \includegraphics[width=3in]{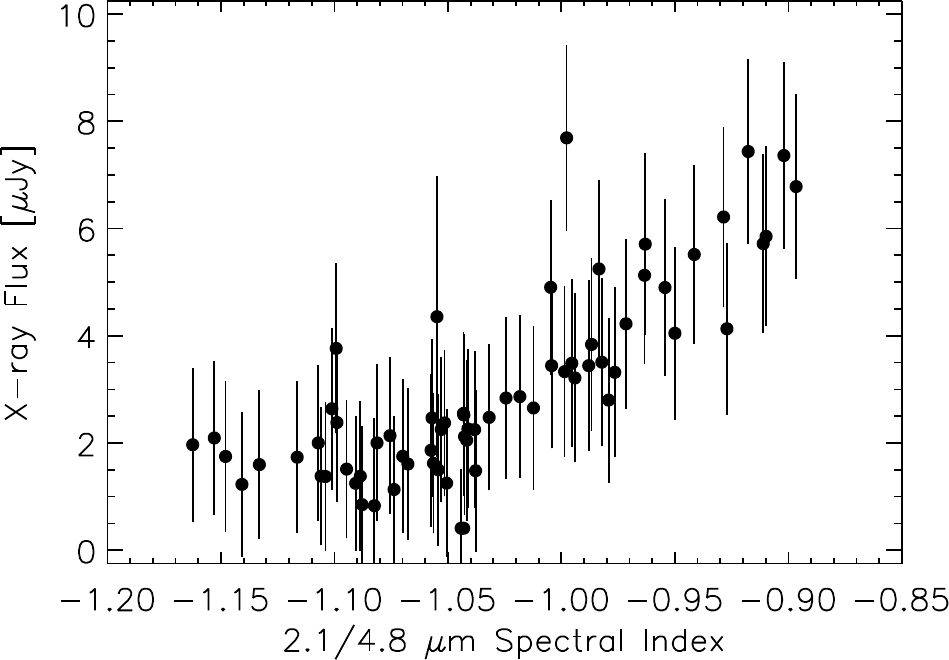} 
   \caption{
{\it (a) Top Left:}
 Correlation of  X-ray flare emission detected by NuSTAR with  NIR flare emission at 2.1 $\mu$m, 
illustrating their similar origin. 
{\it (b) Top Right:}
Similar to (a), except at 4.8 $\mu$m. 
{\it (c) Bottom:}
A plot of the spectral index of NIR flare emission 
against  the X-ray flux density. 
A trend is noted in that  the NIR spectral  index becomes shallower 
as  the  X-ray flux increases. 
The colored line is a linear fit to the full data set, although the correlation seems to cease at $\gtrsim$ 9 and $\gtrsim$ 20 mJy in the IR fluxes.
}
\end{figure}

To examine whether the brightness of X-ray flare emission is correlated with NIR flare emission, Fig.\, 3a and 3b illustrate such a trend between NIR flux density at 
2.1 or 4.8 $\mu$m, respectively, with X-ray flux density. The lines shown in this figure suggest that the NIR and X-ray emission are correlated at fainter levels, as 
was noted in Figures 2b,c  
but that the correlation breaks down at bright levels, where the X-ray goes up without any further increase in the NIR brightness. 
Figures 3a,b shows this  linear trend until the flux reaches to 9 and 20  mJy at 2.1 and 4.8 $\mu$m, respectively, 
where the IR (mostly) stops increasing but the X-ray does get brighter.
To examine the variation of NIR 
spectral index with X-ray emission, Fig.\ 3c shows that the spectral index of NIR flare emission tends to become shallower when the X-ray flux density becomes 
stronger. As discussed above, the spectral index of these brightest flares gets flatter with brightness. 



\begin{figure}[htp] 
   \centering
   \includegraphics[width=3in]{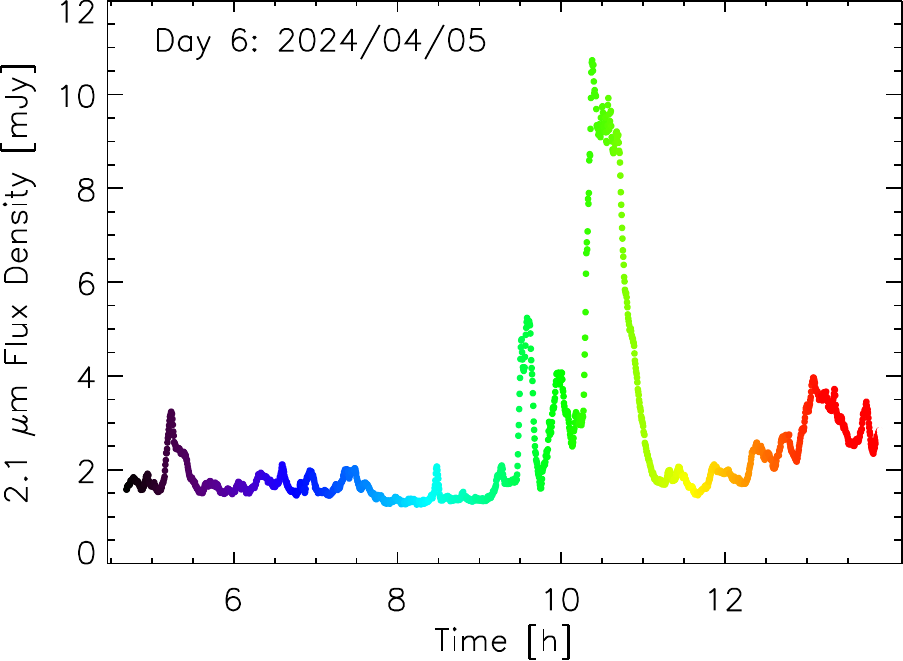} 
   \includegraphics[width=3in]{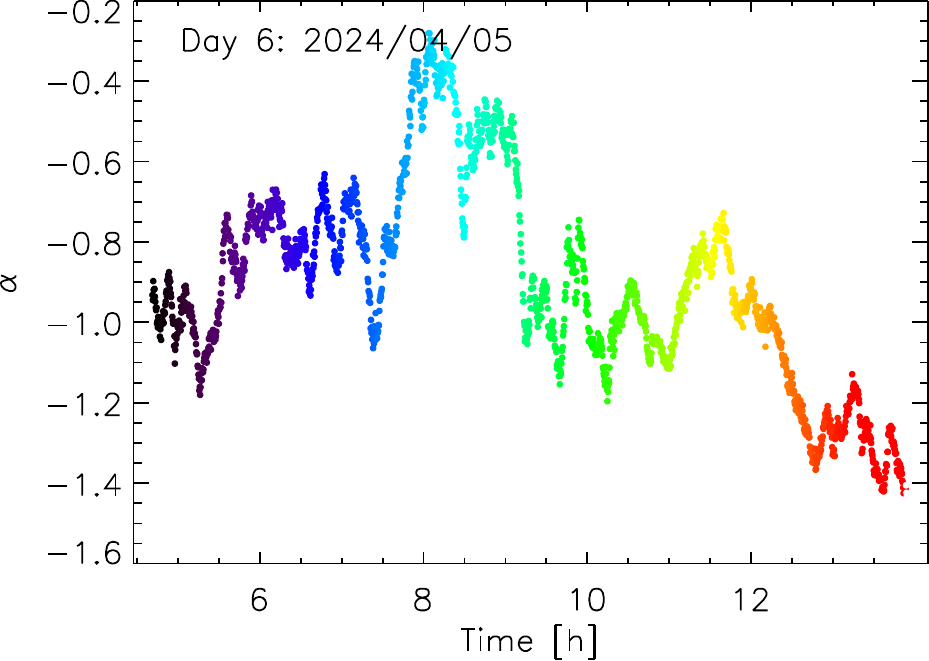} 
  \includegraphics[width=3in]{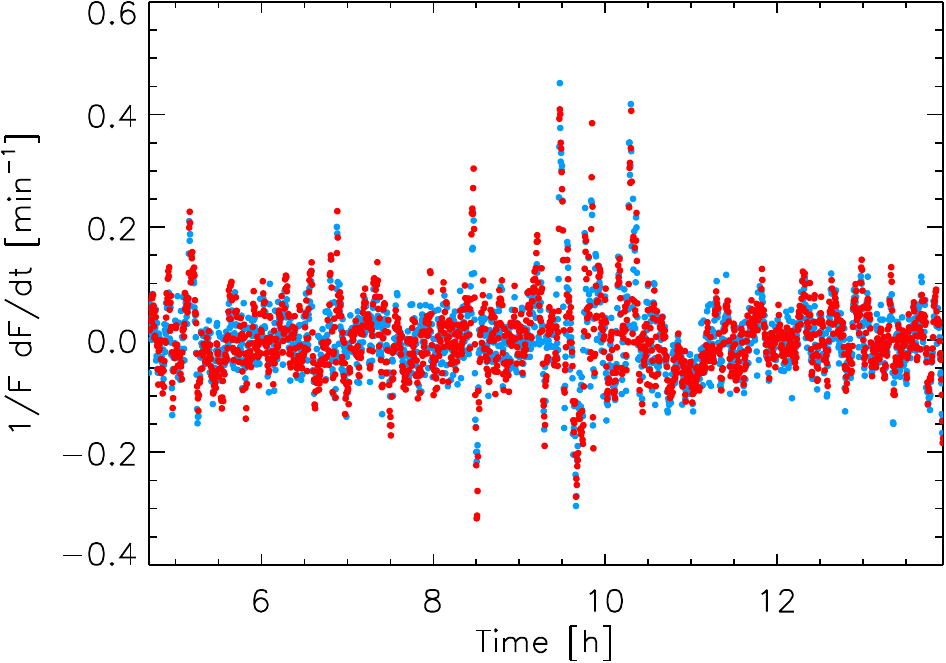}
   \caption{
{\it (a) Top Left:}
Light curve of Sgr A*  at 2.1 $\mu$m taken on April 5, 2024.
{\it (b) Top Right:}
Same as (a), except the NIR spectral index of Sgr A* as a function of time. 
{\it (c) Bottom:}
Same as  (a), except that normalized values of the time derivative of the flux at 2.1 (blue) and 4.8$\mu$m (red) are 
displayed as a function of time. 
}
\end{figure}

For a more detailed plot of a trend in which the population of bright NIR flares show shallow spectral index values, Figs. 4a and b show the light curve and the 
corresponding spectral index between 2.1 and 4.8 $\mu$m \citep{zadeh25}. A correlation of the spectral index with NIR brightness is noted. 
This is evident during the IR flare between 10-11h. 
Though anti-correlation can be seen for the fainter flare at 5-5.5 h, and for the flaring emission at $\ge$12 h.
The spectral index of flickering features after 11h UT, is steeper than bright flares between 10-11h UT.  There are times that this trend for weakest 
features is  not followed, as shown  between 8-9h UT. 
This is because the low flux is very close to the pedestal level, so giving a spurious  spectral index due to 
small errors in the subtraction pedestal level  which can lead to large systematic errors in the spectral index.

What distinguishes flares from flickers is that their brightnesses increase in a strikingly short time. Figure 4c displays the 
flux-normalized derivative of the flux density at 2.1 and 4.8 $\mu$m as a function of time. The sharp spikes indicate sudden rise of the 
flux density over a short time interval. The mean gradients roughly range between $\sim0.1\, \rm{min}^{-1}$, during low-level intensity 
flickering, and $\sim 0.3-0.4\, \rm{min}^{-1}$ for typical flares.  The highest values are noted where the X-ray flare is detected near 
10.35 UT with a gradient $\sim0.45\, \rm{min}^{-1}$, which corresponds to a factor of 2 increase in flux density in 1.5 min.  This short 
time scale places a constraint on the particle acceleration mechanism during flare activity at NIR wavelengths.

\begin{figure}[htp] 
   \centering
   \includegraphics[width=3in]{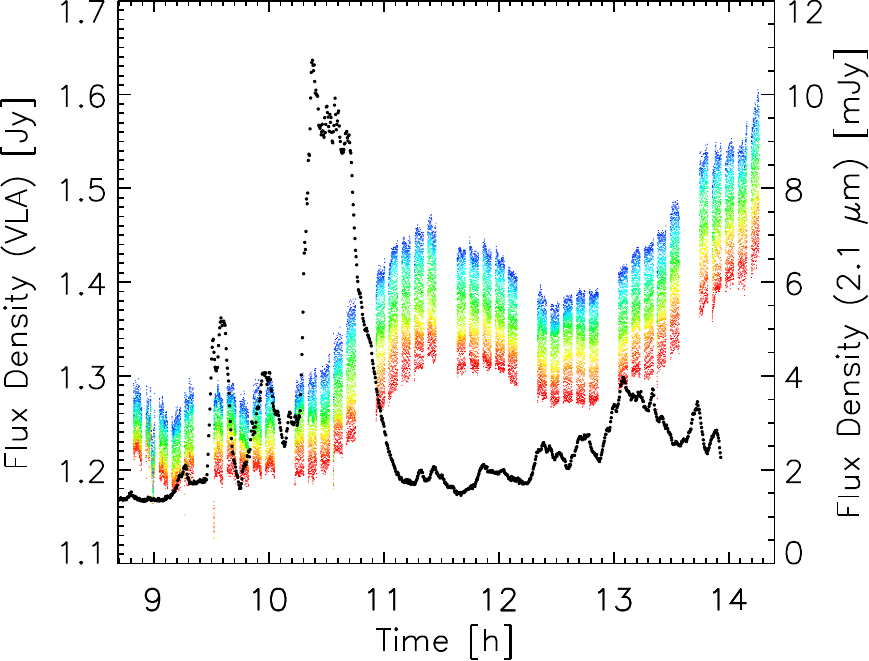}
   \includegraphics[width=3in]{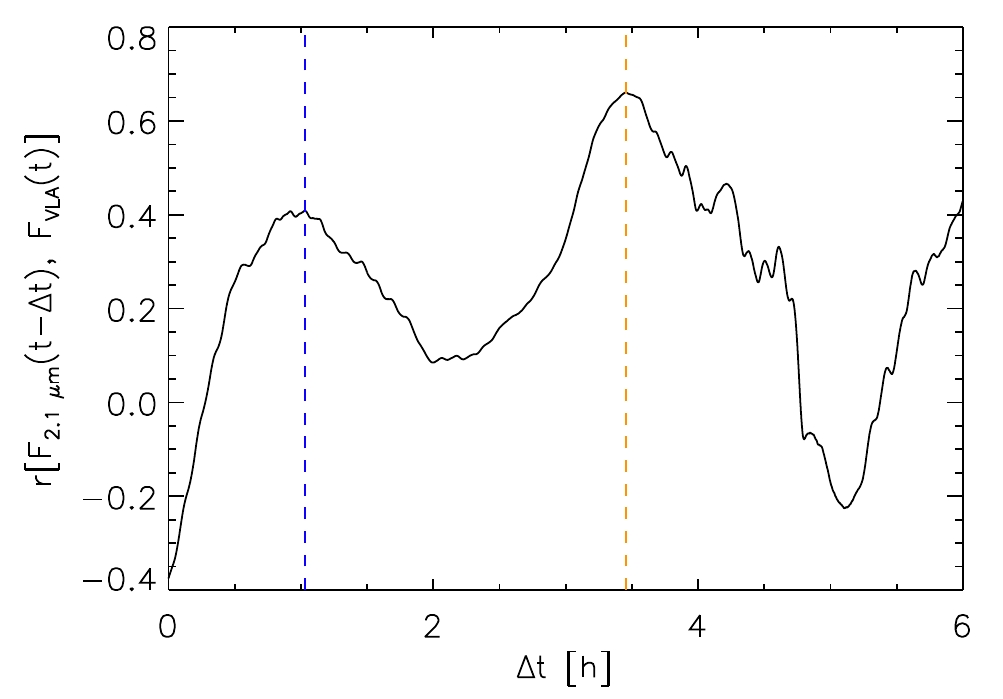} 
\caption{
{\it (a) Left:}
Superimposed on the NIR 
light curve at 2.1 $\mu$m are 
the light curves of Sgr A* at 1 cm, ranging from frequencies 
29   to 37 GHz with colors from red to  blue to violet, respectively. 
Note that the scale for the radio flux density does not start at zero  
(i.e. the contrast in the variation of the radio emission is much lower than in the NIR.)
{\it (b) Right:}
The cross correlation coefficient between NIR and radio as a function of lag time.
There are two peaks shown with dotted lines at  1 and 3.5 h lag times. 
}
\end{figure}

\vfill\eject

\subsection{NIR vs Radio  light curves}
\subsubsection{Time delays}
Another relationship that we investigate is the correlation between the flaring emission in the NIR and radio, which is thought to arise 
from a synchrotron hotspot orbiting within the accretion flow.  The synchrotron source is optically thin in the NIR, with the rise and decay 
of the light curve directly tracking the acceleration of energetic electrons and their subsequent synchrotron losses.  At radio wavelengths 
the source is initially optically thick, and the synchrotron loss time at the relevant electron energies is longer than the 1-2h flare 
duration, suggesting that the variability is due to adiabatic cooling of the electron population as the hotspot expands.  In this picture, 
the flux rises as the optically thick hotspot expands, then falls as it becomes optically thin.  The strongly increasing synchrotron optical 
depth with increasing wavelength implies that the flaring peaks at successively later times at longer wavelengths 
\citep{vanderlaan66,zadeh06a}.

Thus, in this scenario one expects a delay between the NIR and radio flaring and a systematic delay in the flaring across the bandwidth of 
the VLA observations.  In fact, GRMHD simulations in combination with relativistic ray tracing show that flare events due to reconnection of 
the magnetic field produce time lags at lower frequencies due to self-absorption in the accretion disk \citep{jiang25}.

Fig.~5a shows the light curves of Sgr A* at a range of frequencies between 29 and 37 GHz, superimposed on the NIR light curve at 2.1 $\mu$m.  The 
strongest variable emission in the radio is detected at $\sim$ 37 GHz, displayed in red.  The strongest radio emission is detected toward the end of 
the observation near 14.2h UT. There is also a (weaker) peak in the radio emission near $\sim$11:30 UT. At NIR, the strong, primary peak is detected 
between 10 and 11h UT, whereas a cluster of secondary peaks are between 9.5 to 10h UT. The cross correlation of NIR and radio, as displayed in Fig.\ 
5b, shows two (primary and secondary)  correlation peaks.  {We note that the peaks in the primary and secondary correlation indicate time delays 
of 3.5 or 1 hour, respectively. The time delays are selected with respect to the brightest NIR flare which peaks between 10:30h to 11:00h UT.  The 
primary time delay at 3.5h is a lower limit because the radio brightness may not have peaked before the observations ended near 14:15 UT.}

\begin{figure}[htp] 
   \centering
   \includegraphics[width=3in,angle=0]{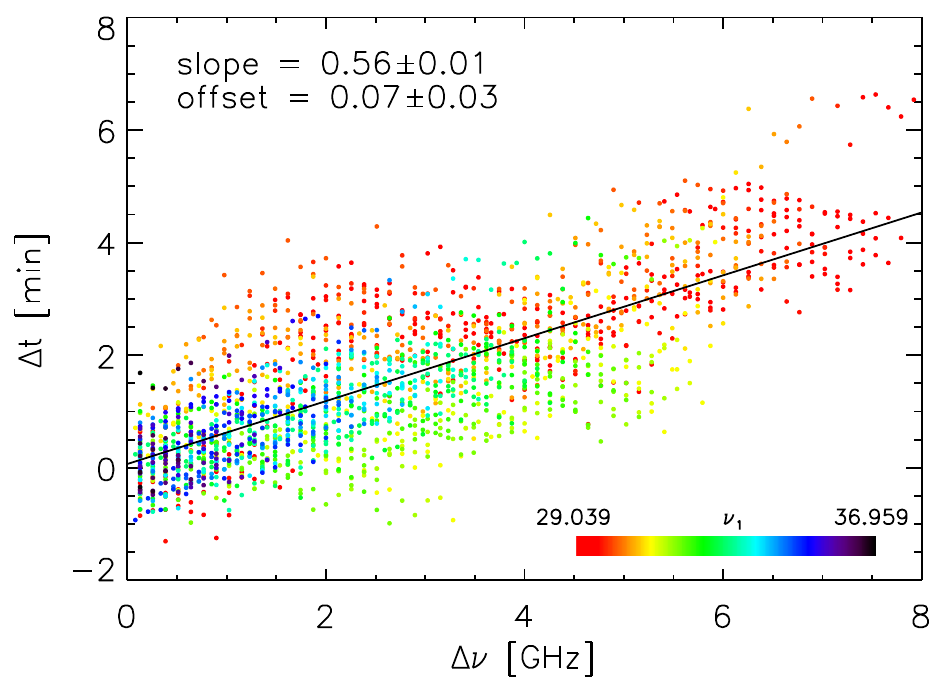} 
   \includegraphics[width=3in,angle=0]{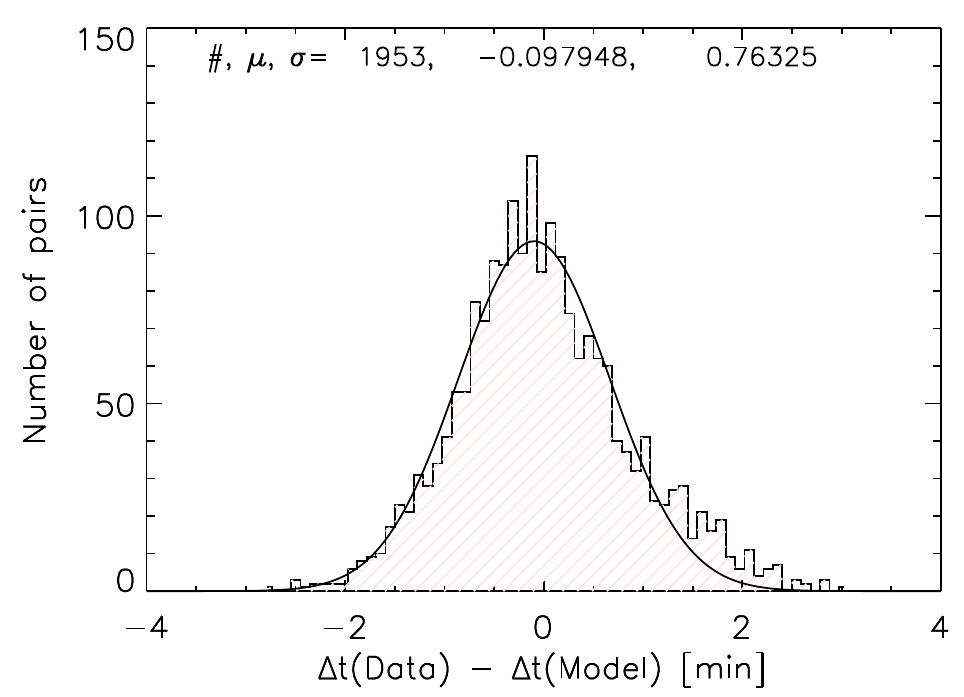} 
   \caption{
{\it (a) Left:}
Calculated time delay between pairs of radio bands, plotted as a function of the
separation between the bands. Points are color coded according to the frequency 
of the low frequency band ($\nu_1$) in each pair. A linear fit to the trend 
indicates a mean delay of 4.5 min for bands separated by 8 GHz.
{\it (b) Right:}
Histogram of the deviations of the data from the linear model in (a).
The annotation lists the number of data points, and the mean and $\sigma$ of a 
gaussian fit to the histogram, indicating $\sigma = 0.76$ min for the model.
}
\end{figure}

Much of the uncertainty in associating radio and NIR variable emission stems from the limited time ($\sim 6.5$h)  that the VLA can observe the Galactic center.  
However, a more detailed analysis of time delays at radio wavelengths can be made by searching for time delays across the 8 GHz bandwidth. To this end, Fig.\ 6a 
shows time delay vs frequency difference between each pair of the 64 spectral windows, with a linear fit to the delay. 
with a slope of 0.56 min GHz$^{-1}$  and a time delay of 4.5 min.
The data points in the figure are constructed 
by interpolating the light curves at each frequency to a regular grid in time before subtracting the mean slope of each light curve; then the FFT cross-correlation 
between the light curves at each pair of frequencies was calculated. Finally, the offset ($\Delta t$) was determined from the peak of a 4th-degree polynomial fit to 
the peak of the cross correlation (using a range of $ -30 {\rm min} \le\Delta t \le 30 {\rm min}$).  Figure 6 shows the histogram of deviations from the linear fit 
and its Gaussian fit  with a dispersion of 0.76 min.

\begin{figure}[htp] 
  \centering
   \includegraphics[width=3in]{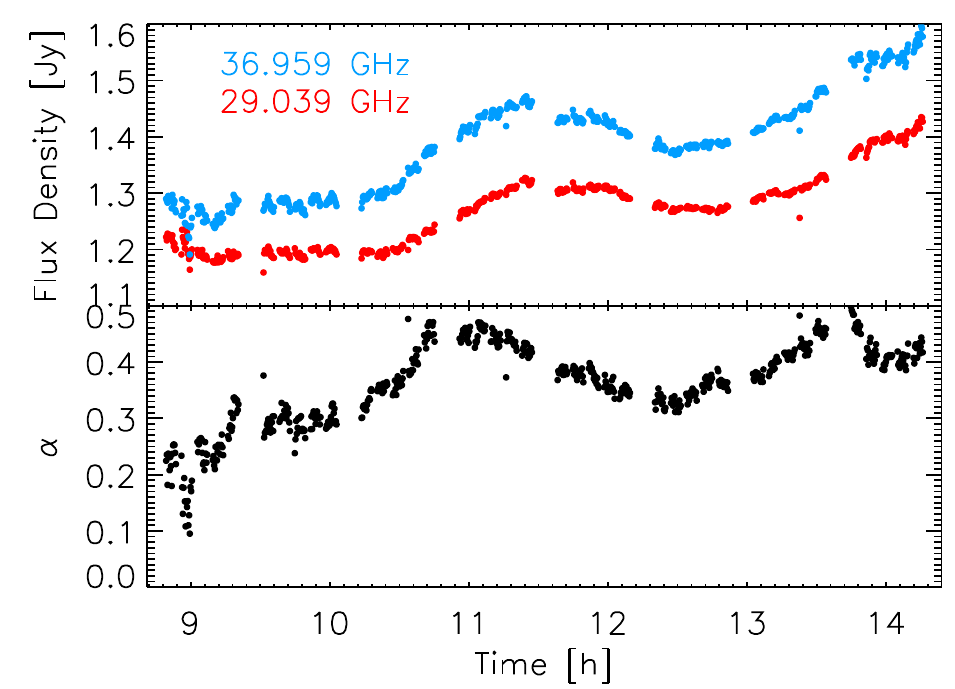} 
  \includegraphics[width=3in]{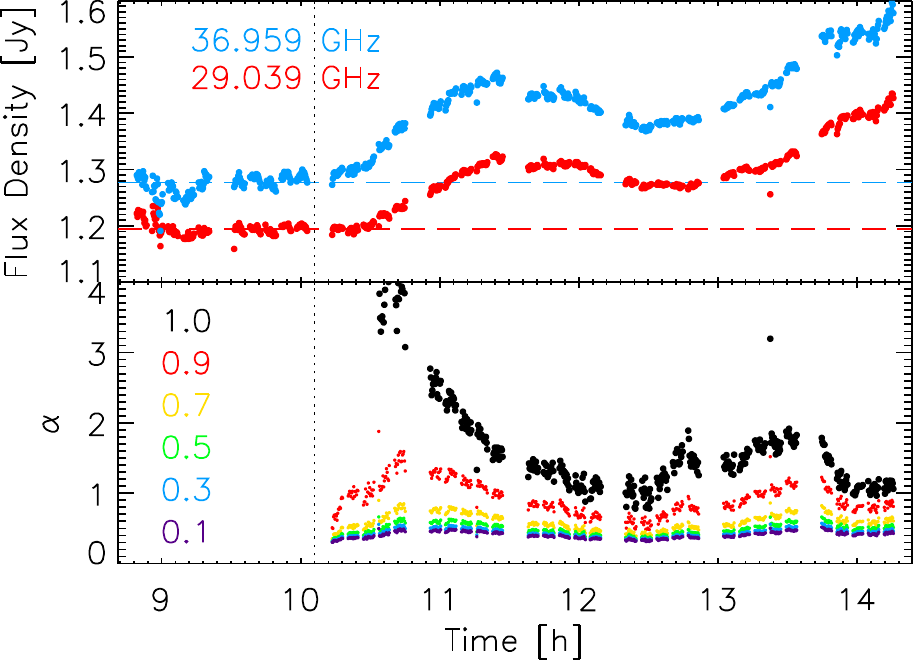} 
  \caption{
{\it (a) Left:} 
The plots show the light curves at the lowest and highest radio frequencies, and 
the total spectral index, as a function of time {\bf (black dots)}  when the variable and quiescent flux densities  are combined. 
{\it (b) Right:} 
{The plots show the spectral index as a function of time, calculated after subtraction of the median flux density for times $<$ 10.1h UT. The 
time limit is indicated by the vertical line, and the subtracted flux densities are indicated in the upper panel.
Additional spectral index results {(colored dots)} are also shown if the backgrounds  in the top panel are reduced 
by the factors ranging between 0.9 and 0.1, as listed.} 
}
\end{figure}

\subsubsection{The Spectral Index}

In order to measure the evolution of the spectral index with time, Fig.\ 7a shows the spectral index between the lowest and highest 
frequencies.  The values of $\alpha$ range between 0.2 to 0.5, with a mean value of $\sim$0.35. In Fig.\ 7b, after subtracting a constant 
component of the emission, the spectral index of the remaining variable emission shows a decline before the flux density peaks at 
$t\sim$11.5h. After this peak, the spectral index is loosely anticorrelated with the flux density. 
In  a picture of an 
optically thick expanding hot plasma, the opacity drops as the plasma expands. The very positive spectral index in the radio is due to 
optically thick synchrotron emission, which is unlike the NIR synchrotron emission where the spectral index is negative, tracing optically 
thin emission. However, there is uncertainty due to the  background level  in measuring the spectral index. 
{If the backgrounds shown in the top panel are reduced
 by the factors listed to the left in  Fig.\ 7b, alternative  spectral index values are presented. 
At a factor of 0.1 (dark purple) the spectral index is very close to that shown in Figure 7a. 
This shows  that the spectral index is sensitive to the background that is subtracted when that background is a 
large fraction of the observed signal.}

\subsection{Analysis of Radio Variability}
 
{The restricted overlap of the VLA and JWST light curves and the lack of sub-mm observations able to bridge the large frequency range between 
them suggests a weak connection the radio and NIR light flaring. Here, we focus on the implications of the lag between the low and high radio 
frequencies for models of the radio emission.}  


The  level of the underlying ``quiescent'' emission clearly changes during the flare and the small separation in frequencies means that the frequency-dependence of the peak flares and profile widths are poorly constrained.   To illustrate these difficulties we contrast two expanding synchrotron source models for the flaring emission.   A  ``thermal'' model adopts a relativistic Maxwellian electron spectrum, which for electron temperature $T_e$ takes the form $E^{-2} \exp(-E/kT_e)$.  An alternative  ``non-thermal'' model adopts a power-law electron population with an $E^{-p}$ energy spectrum.  In both cases the source is assumed to be a homogeneous sphere with radius $R(t)$ that increases linearly with time.  To minimize the number of parameters we assume that the source expands with constant speed, i.e.  that
\begin{equation}
    R/R_0=1 + (t-t_0)/t_\mathrm{exp}
    \label{eqn:RoR0}
\end{equation}
where $t_0$ is the time of the peak in the flux at frequency $\nu_0$ and $t_\mathrm{exp}$ is the expansion time scale. 
The magnetic field threading the source 
and number density of synchrotron-emitting electrons declines as $R^{-2}$ and $R^{-3}$, respectively.   
In addition,  the emitting electrons undergo adiabatic cooling, 
their individual energies declining as $R^{-1}$ as the source expands.

{
\subsubsection{Thermal model}
The received flux is 
\begin{equation}
    S_\nu = 
 S_0 \, \left(\frac{\nu}{\nu_0}\right)^{\!\!2} \, \frac{R}{R_0} \; \frac{1-e^{-\tau_\nu}}{1-e^{-\tau_0}} \,.
 \label{eqn:Snu-flare-t}
\end{equation}
where $S_0$ is the peak flux attained at frequency $\nu_0$, and $R_0 = R(t_0)$ is the source radius at time $t_0$ when this occurs.  The 
optical depth of the source is \footnote{For simplicity, the escape probability from a sphere is approximated by 
$(1-e^{-\tau_\nu})/\tau_\nu$, where $\tau_\nu$ is 4/3 times the optical depth from center
 to surface. This is exact in the large and small optical depth limits with less than 5.2\% error in-between.}
\begin{equation}
    \tau_\nu= \tau_0 \left(\frac{\nu}{\nu_0}\right)^{\!\!-1} \, \frac{R}{R_0}\,
    \frac{M(x)}{M(x_0)}    \,.
    \label{eqn:taunu-flare-t}
\end{equation}
Here
\begin{equation}
M(x) = 4.0505\,x^{-1/6}\left( 1 + 0.4\,x^{-1/4} + 0.5316\,x^{-1/2}\right)
 \exp\left(-1.8899\,x^{1/3}\right) 
    \label{eqn:Mx}
\end{equation}
is a fitting function for thermal synchrotron emissivity \citep{mahadevan96}, and 
\begin{equation}
    x = \frac{4\pi m_ec\nu}{3eB} \left(\frac{m_e c^2}{kT_e}\right)^{\!\!2} = x_0 \,\frac{\nu}{\nu_0} \, \left(\frac{R}{R_0}\right)^{\!\!4}\,.
    \label{eqn:x-t}
\end{equation}
To guarantee that $S_{\nu_0}$ peaks when $R=R_0$,  $\tau_0$ must satisfy
\begin{equation}
    e^{\tau_0} + \left(1+4\,x_0 \frac{M'(x_0)}{M(x_0)}\right)\,\tau_0 - 1 = 0 \,.
    \label{eqn:tau0-t}
\end{equation}
where $M' = dM/dx$.}

\subsubsection{Non-thermal model}
In this case the received flux is
\begin{equation}
    S_\nu= 
    S_0 \, \left(\frac{\nu}{\nu_0}\right)^{\!\!5/2} \, \left(\frac{R}{R_0}\right)^{\!\!3} \; \frac{1-e^{-\tau_\nu}}{1-e^{-\tau_0}} \,,
 \label{eqn:Snu-flare-nt}
\end{equation}
and the optical depth of the source is
\begin{equation}
    \tau_\nu = \tau_0 \left(\frac{\nu}{\nu_0}\right)^{\!\!(p-4)/2} \, \left(\frac{R}{R_0}\right)^{\!\!-(3+2p)}\,,
    \label{eqn:taunu-flare-nt}
\end{equation}
\citep{vanderlaan66}, where $\tau_0$ must satisfy
\begin{equation}
    e^{\tau_0} - (1+2p/3)\,\tau_0 - 1 = 0 \,.
    \label{eqn:tau0-nt}
\end{equation}
to guarantee that $S_{\nu_0}$ peaks when $R=R_0$   \citep{zadeh06a}.

{ 
\subsubsection{Underlying emission}
In addition we require a model for the slowly-varying``quiescent'' emission that underlies the flaring component. As this appears to be steady prior to the flare's commencement and rising afterward we adopt  a simple two-component model that is steady until time $t_b$, then rises linearly at rate $b$:
\begin{equation}
    S_{\nu,q}(t) \;=\;  a_\nu + u(t-t_b)\, b_\nu \,
 \label{eqn:Snu-quiescent} 
\end{equation}
where $u(t-t_b)=0$ for $t<t_b$ or $u(t-t_b)=t-t_b$ otherwise.   Note that $t_b$ is the same at all frequencies.}

\subsubsection{Fitting to radio light curves}

{The thermal and non-thermal flare models both have 4 free parameters.  Three of these are in common: $S_0$ and $t_0$, the magnitude and timing of the 
peak flux at frequency $\nu_0=36.831\,GHz$, and the expansion time scale $t_\mathrm{exp}$. The fourth parameter characterizes the form of the 
electron energy spectrum, temperature parameter $x_0$ or the power-law index $p$ for the thermal and non-thermal models, respectively.  In addition, 
the underlying slowly-varying component has 5 free parameters: the constants $a_{\nu_0},\, a_{\nu_1}$, the time $t_b$ when the linear rise begins, 
and the subsequent rates of increase $b_{\nu_0},\,b_{\nu_1}$.

{
We perform a simultaneous maximum likelihood fit of each combined quiescent-plus-flare model to the 29.103 and 36.831 GHz light curves.  The fit is not weighted by the measurement errors because 
they are dominated by short time scale fluctuations.  To estimate the constraints on the model parameters we use the rms deviation of the points from the model as an estimate of the effective 
$\sigma$ for the data.  The fitted parameter values and their formal statistical errors for the non-thermal and thermal models are listed in Table \ref{tab:F8_radio-fit}, and the best-fit models 
are over-plotted on the observed light curves in Figure 8.}
\begin{figure}
\plottwo{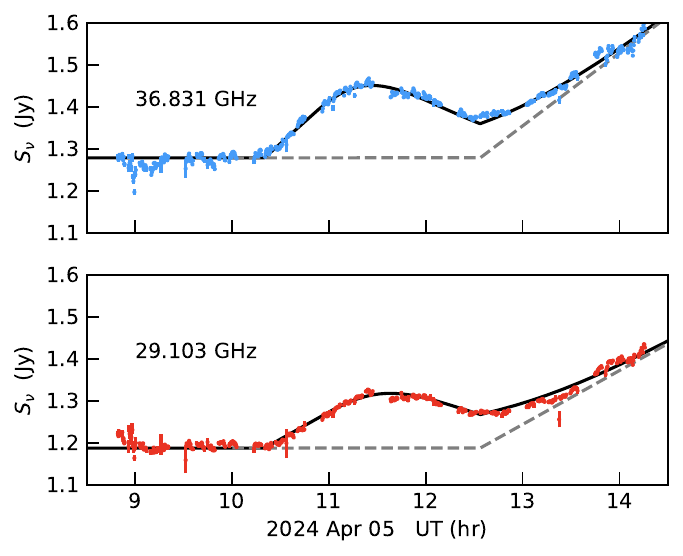}{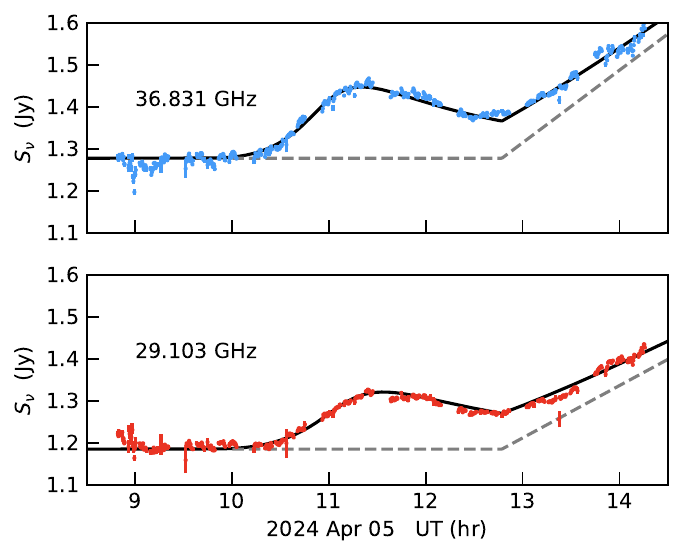}
\caption{
{\it (a) Left:} 
{ Best-fit adiabatically-expanding synchrotron source models for the radio light curves on 2024 April 05, assuming  a relativistic 
Maxwellian distribution of electron energies or
{\it (b) Right:}  a power-law energy spectrum.  The observed light curves at 36.8 and 29.1 GHz are 
plotted in blue and red, respectively.  Gray dashed lines indicate the underlying slow variation of the quiescent flux, solid curves include 
the addition of a flaring component contributed by an expanding synchrotron source which peaks as it transitions from being optically thick 
to thin (see text).}}
\label{F8_radio-fit}
\end{figure}
\begin{table}[htp]
\caption{Model parameters and formal errors for best-fit radio light curves}
\begin{center}
\begin{tabular}{l @{\qquad} r c l @{\qquad} r c l }
&  \multicolumn{3}{c}{Non-thermal} & \multicolumn{3}{c}{Thermal} \\
$S_0$\,(Jy)  & 0.1696 & $\pm$ & 0.0014	& 0.1726 & $\pm$ & 0.0014 \\
$p$,\quad$x_0$  & 0.81 & $\pm$ & 0.03 		& 0.175 & $\pm$ & 0.022 \\
$t_0$\,(hr) & 11.343 & $\pm$ & 0.008 	& 11.442& $\pm$ & 0.009 \\
$t_\mathrm{exp}$\,(hr) & 1.69 & $\pm$ & 0.04 & 1.107 & $\pm$ &  0.015\\
$a_0$\,(Jy) & 1.2778 & $\pm$ & 0.0009 	& 1.2786 & $\pm$ & 0.0008\\
$a_1$\,(Jy)  & 1.1852  & $\pm$ & 0.0007 	& 1.1876 & $\pm$ & 0.0007\\
$t_b$\,(hr)   &  12.786 & $\pm$ & 0.012 	& 12.561 & $\pm$ & 0.014\\
$b_0$\,(Jy/hr) &  0.1727 & $\pm$ & 0.0019 & 0.1737 & $\pm$ & 0.0016\\
$b_1$\,(Jy/hr) & 0.1247 & $\pm$ & 0.0016 	& 0.1284 & $\pm$ &0.0013\\
\end{tabular}
\end{center}
\label{tab:F8_radio-fit}
\end{table}
{The thermal and non-thermal models are both able to reproduce the observed light curves, as shown in Fog.\ 8, and the small formal errors on their parameters indicate that the models are 
tightly constrained, i.e. that adjustments to any parameters outside the formal error range significantly increases $\chi^2$.  The 5 baseline parameters and $S_0$, $t_0$ are directly related to the 
features of the light curve, leaving just $t_\mathrm{exp}$ and either $p$ or $x_0$ as available to probe the physical differences between the models. In the parametric view, the difference between 
the models is that the non-thermal one has a more slowly decaying tail than the thermal model. This allows $t_b$  to be larger for the non-thermal model since less background is needed at $t \ge t_b$.}} 


{The data alone are not able to distinguish between the two models, and one can imagine that other electron energy spectra would also provide a reasonable fit.  However, we can gain a little more 
insight by taking each fit at face value and deriving illustrative physical conditions in the source.  To do this we start with the values of $S_0$, $\tau_0$ and either $x_0$ or $p$ (for the 
thermal and non-thermal models, respectively) and assume energy equipartition with the magnetic field.}

{For the thermal model we assume that the emitting plasma has an equal number density of protons with $T_i = 10 T_e$, in line with the standard picture of the accretion flow onto Sgr A*.  Assuming 
energy equipartition with the magnetic field, we find $B_0 \approx 45\,$G, $n_e\approx 9.0\times10^4$\,cm$^{-3}$, $kT_e \approx 17\,$MeV and $R_0 \approx 3.8\,r_g$.  If, however, the magnetic 
energy density is reduced to 10\% of the plasma energy density we obtain $B_0 \approx 26\,$G, $n_e\approx 2.3\times10^5$\,cm$^{-3}$, $kT_e \approx 22\,$MeV and $R_0 \approx 3.2\,r_g$.  These 
numbers, while uncertain, are broadly in line with what is expected in the Sgr A* accretion flow (e.g., see \S5.3) and so are consistent with the idea that the radio variability arises from 
transient inhomogeneities in the accretion flow.}


{For the non-thermal model the unusually flat $E^{-0.8}$ electron spectrum means that the equipartition estimate is sensitive to the upper cutoff in 
the electron spectrum.  Assuming that the power-law electron spectrum extends between 5 and 100\,MeV and equipartition with the magnetic field, then 
$R_0 \approx 4.4\,r_g$ and $B_0 \approx 11\,$G, increasing to $4.6\,r_g$ and 13\,G if the cutoff is increased to 200\,MeV.  The energy of the 
electrons radiating most strongly at 36\,GHz is only 15\,MeV, similar to the temperature in the inner accretion flow, undercutting the notion that 
the radio flaring might arise from the late evolutionary stages of the source responsible for earlier flaring in the IR/X-rays.

To summarize, while the models are by no means unique, a simple thermal model can match the radio light curves assuming physical conditions 
similar to those inferred for the inner accretion flow (see \S5.3)  suggests that occasional flaring in the radio is a byproduct 
of the chaotic accretion flow, as seen in simulations \citep[e.g.][]{grigorian24, jiang25}.}

\section{Models of X-ray flare emission}

Here we examine the viability of synchrotron and inverse Compton models to explain the X-ray flaring and how it ties in to NIR flare emission.
We represent the accretion flow for simplicity  as a ``disk'' emitting steady synchrotron radiation primarily in the submm, plus a transient compact flare source emitting synchrotron in the NIR.  

\subsection{Synchrotron Mechanism}

Synchrotron emission models for X-ray flaring are often invoked because the X-ray flares have spectral indices consistent with synchrotron cooling break models \citep{dodds-eden09}, although uncertain 
because of the significant correction for interstellar absorption. Either the X-ray and NIR spectral indices are consistent, or the X-ray spectral index is steeper, and the flux is consistent with a 
synchrotron cooling break, \citep[e.g.][]{ponti17,haggard19}.  
One generic difficulty  is that synchrotron losses for the high-energy electrons emitting in X-rays are severe, with loss times of $\sim 1$ second for 
$B\sim100$\,G. 

\begin{figure}
    \centering
\includegraphics[width=9.5cm]{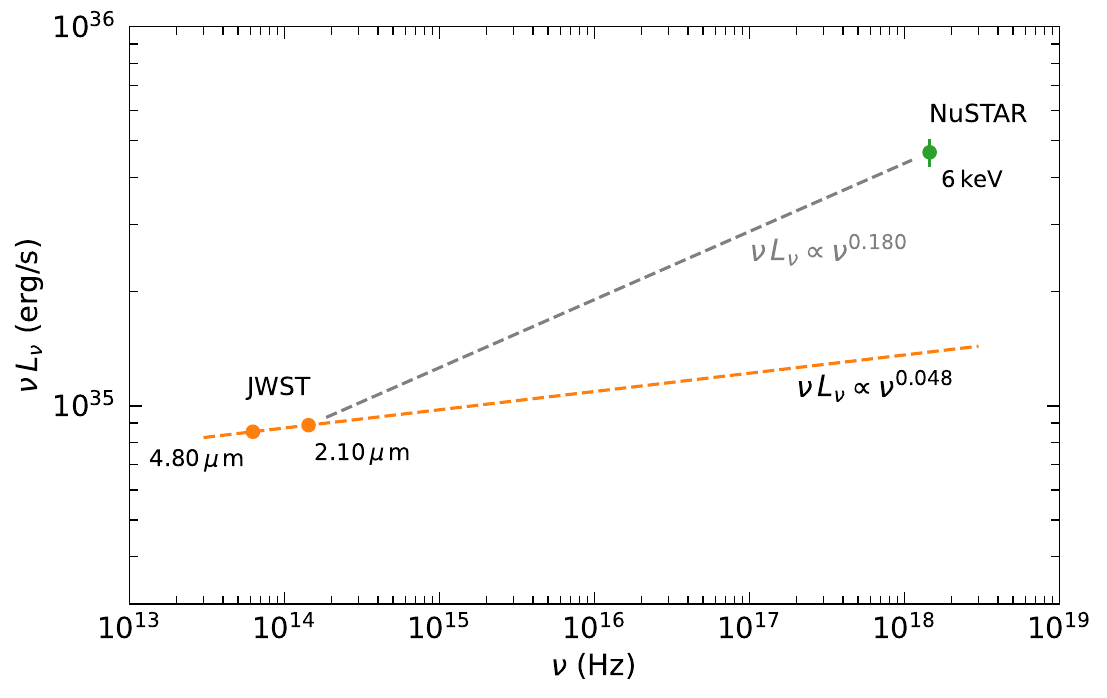}
\caption{
The infrared to X-ray spectral energy distribution at the peak of the flare at UT 10.5h on 2024 Apr 05.  Orange and green points show the JWST and NuSTAR peak luminosities at 
4.8\,$\mu$m, 2.1\,$\mu$m, and 6\,keV, respectively.  The measurement errors of the JWST measurements are at the 1\% level, within the size of the data points. Dashed lines show the 
power-law extrapolation of the IR spectral index (orange) and the harder power-law spectrum that would connect the 2.1 and 6\,keV measurements (gray).
\label{fig:IR-Xray-spectrum}
}
\end{figure}

The NuSTAR flare, combined with NIRCam data taken at two different wavelengths, represents a challenge for synchrotron models because it is particularly bright 
relative to the NIR.  In Figure \ref{fig:IR-Xray-spectrum} we compare the peak luminosities at 4.8 and 2.1\,$\mu$m obtained with JWST with the NuSTAR peak flux. We 
express the X-ray luminosity at 6\,keV as derived from the deabsorbed 2-10\,keV flux of $7^{+3}_{-2}\times10^{-11}$ ergs cm$^{-2}$ s$^{-1}$ and the photon index, 
with power-law index of 2.1$\pm0.3$.  We note that the NuSTAR flux is about 3.5 times the value predicted from power-law extrapolation from the two NIR measurements. 
This requires a harder spectral component at higher energies, which is difficult to reconcile with the short synchrotron loss time for electrons emitting in the 
X-ray.  Overall, this suggests that synchrotron emission is not responsible for the X-ray flaring.

Another argument that can be made against the viability of synchrotron mechanism is from NIR flare emission which is known to be due to optically thin synchrotron radiation. The 
spectral index of NIR flares taken with NIRCam show evidence of synchrotron cooling between 2.1 and 4.8 $\mu$m \citep{zadeh25}. The spectral index of NIR emission between 2.1 and 4.8 
$\mu$m showed a steepening of the spectrum in all seven epochs of JWST observations. This steepening was best shown as flare emission that was decaying with time.  This is also 
supported by cross correlation of 2.1 and 4.8 $\mu$m data, indicating that the 2.1 $\mu$m emission was delayed by 10 seconds on April 5, 2024. The 10 second time delay was modeled as 
due to synchrotron cooling. If the X-ray flux is indeed due to synchrotron emission and the same population of particles produces NIR and X-ray flare emission, then continuous 
re-acceleration of particles on $\sim$1 second timescales is in contradiction with the measured NIR synchrotron cooling. These imply that synchrotron X-ray flare emission should have 
shown a cooling break, which is not detected (see Fig.\ 9).

\begin{figure}
    \centering
\includegraphics[width=7cm]{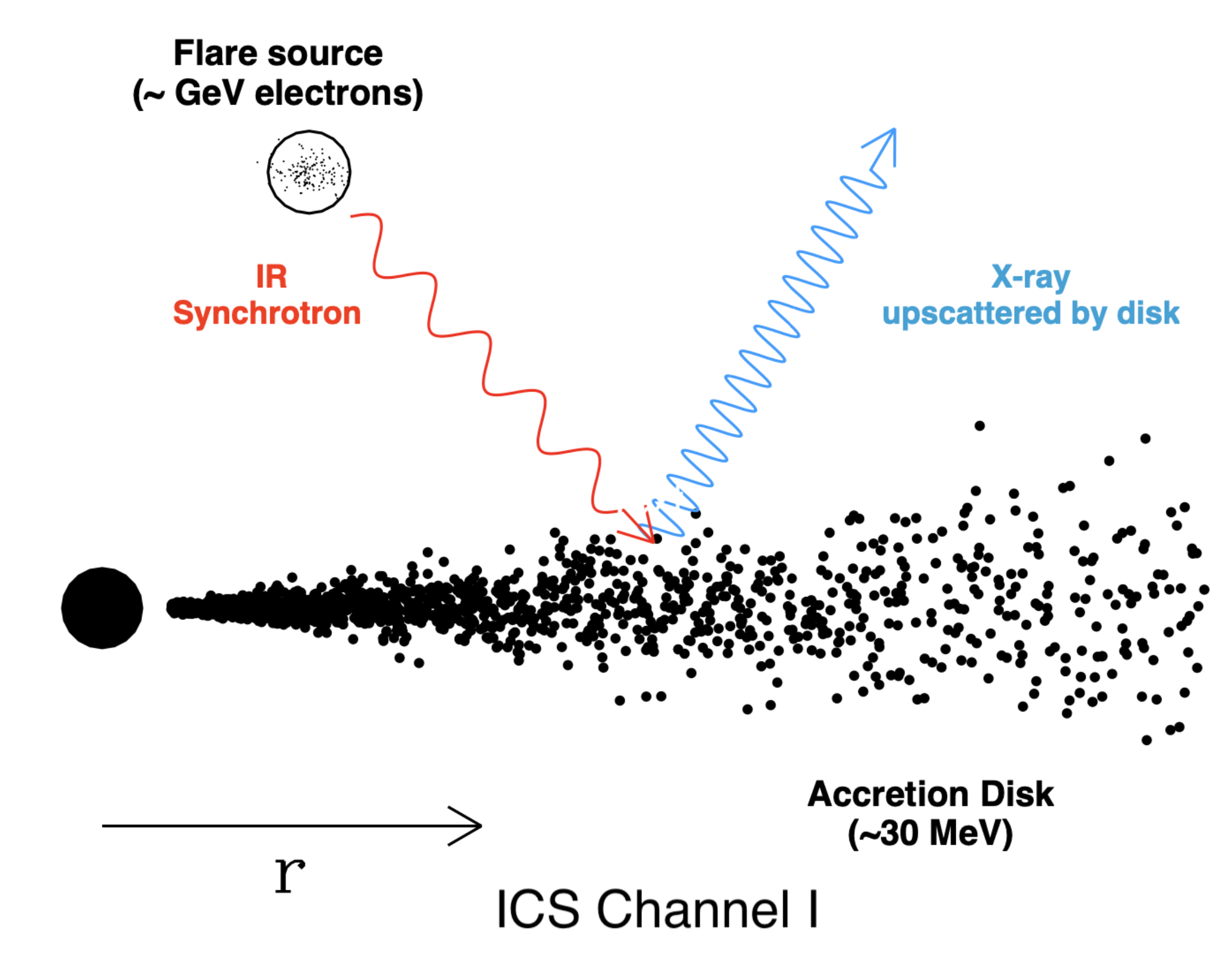}
\vspace{0.2in}
\includegraphics[width=7cm]{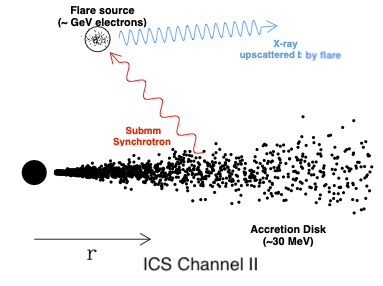}
\vspace{0.20in}
\includegraphics[width=7cm]{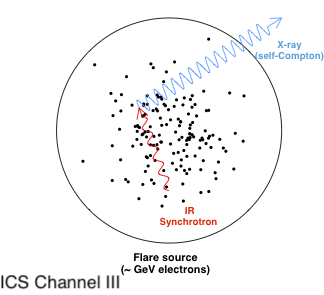}
\vspace{0.2in}
\includegraphics[width=8cm]{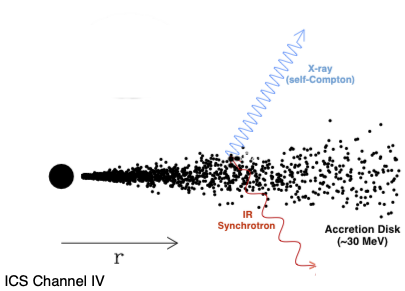}
\caption{
Schematic diagrams of the four different channels of inverse Compton up-scattering that produce X-ray emission from Sgr A*. 
Diagrams a-c (top left, top right and bottom left) 
relate to flaring X-ray emission driven by the transient increase in the electron population in the NIR flare source whereas diagram d (bottom right)
relates to steady background X-ray emission from the accretion disk, as shown in cross-section on one side of  Sgr A*.  
}
\end{figure}

\vfill\eject
\subsection{Inverse Compton Scattering}

As an alternative, we explore the viability of Compton up-scattering models to explain the X-ray flaring and its relationship to NIR flare emission.  
In this scenario the X-ray photons have been 
up-scattered from the sub-mm or NIR by the relativistic electrons that emit synchrotron emission in those bands.
    
There are two distinct populations of relativistic electrons. A quasi-steady, thermal population with $kT_e \sim10$\,MeV in the thick disk-like accretion flow around Sgr A* which is responsible 
for the quiescent, 
always-present, and somewhat variable, radio to submm emission from Sgr A*.  Flaring in the NIR is produced by synchrotron emission from a transient 
population of non-thermal electrons that are accelerated to 
$\sim$ GeV energies.  For simplicity we refer to these two components as the disk and flare, respectively.

The presence of the two populations implies that there are four channels for Compton up-scattering, as schematically drawn in Figure 10. 
  Three of these produce X-ray flaring:
\begin{enumerate}
\item[1.] NIR synchrotron photons emitted by the flare source are inverse Compton scattered by the electron population in
    the disk (Channel I), 
\item[2.] submm synchrotron photons emitted by the disk are inverse Compton scattered by the electron population in the flare source (Channel II), 
\item[3.] 
NIR synchrotron photons emitted by the flare source are up-scattered by the parent electron population in the flare,
i.e.~synchrotron self-Compton emission from the flare source (Channel III),
\end{enumerate}
A fourth up-scattering channel produces quasi-steady X-ray emission:
\begin{enumerate}
\item[4.] 
synchrotron self-Compton emission from the thermal electron population in the disk (Channel IV).
\end{enumerate}

\begin{figure}
    \centering
\includegraphics[width=9.5cm]{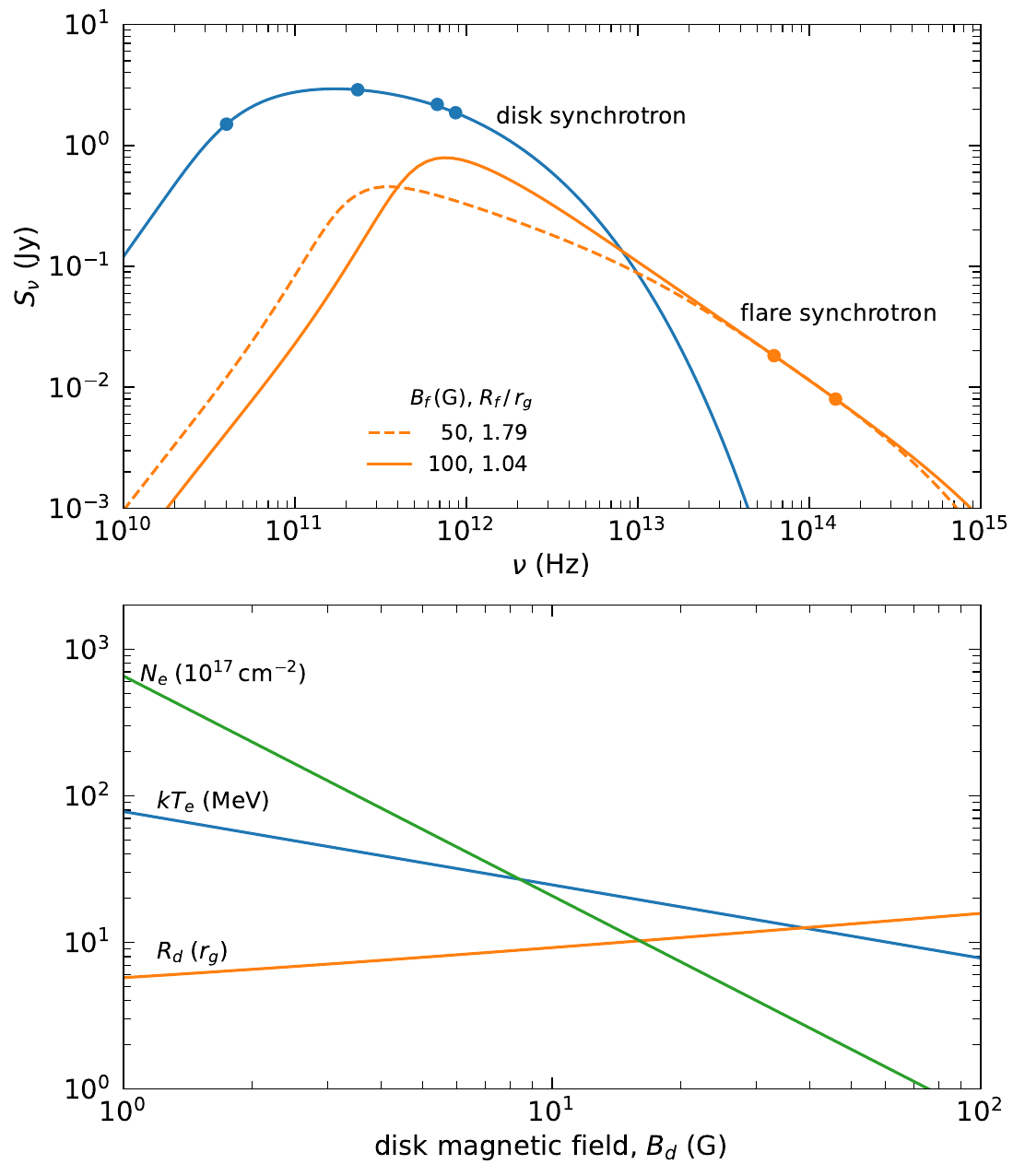}
\includegraphics[width=9.5cm]{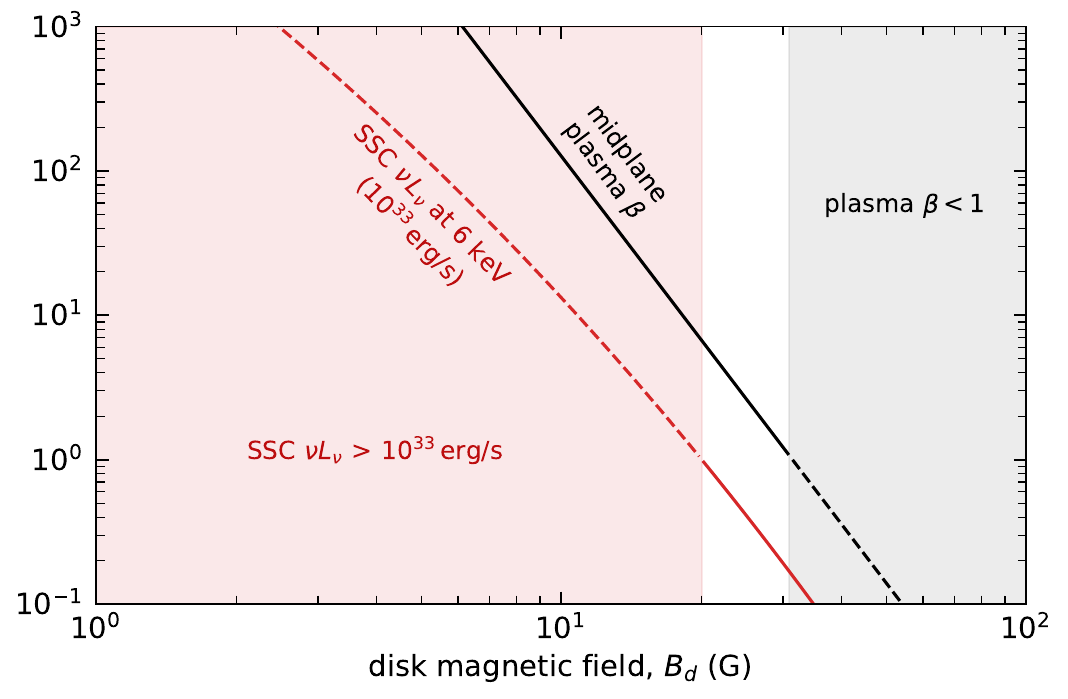}
\caption{
({\it a) (Top:})
The blue curve shows the synchrotron flux density of a homogeneous thermal disk, fit to the observed fluxes at four frequencies from radio to submm (solid 
points).  The orange curves show the synchrotron emission from uniform sphere models for the flaring source region, for different assumed fields and
different radii.  The power-law electron population has been chosen to match the JWST measurements at 2.1 and 4.8 $\mu$m 
with $S_\nu\propto\nu^{-1}$ in the optically thin part of the spectrum.
({\it b) (Middle:})
Disk parameters that produce the blue synchrotron spectrum shown in (a) as a function of disk field strength, $B_d$, radius in gravitational radii 
$r_g$ (blue), electron temperature (red), and electron column density (green).  
({\it c) (Bottom:})
Constraints on the magnetic field of the disk models presented in {\it (a)} and {\it (b)}. 
First, the synchrotron self-Compton X-ray luminosity (maroon curve) must be less than $\sim10^{33}$  erg s$^{-1}$ to be 
consistent with Chandra limits on quasi-steady X-ray emission from Sgr A*, ruling out the region $B_d \le 20$ G (pink shading).  
Second, the gas pressure at the midplane should dominate 
magnetic pressure, i.e. the plasma beta (black curve) should be greater than 1, eliminating the region $B_d \ge\,$  30 G (gray shading).  
We conclude that $B_d$ should lie in the range 20-30 
G (see text). 
}
\end{figure}


To keep free parameters to a minimum,  we adopt simple one-zone models of the disk and flare sources and examine the predicted X-ray fluxes from the 
first three different mechanisms.   Before we examine the viability of ICS mechanism in $\S5.3$ and $\S5.4$, 
we need to determine the electron populations and photons in the disk and the flare source.

%
%
%

\subsection{Disk model}

The accretion flow is modeled as a uniform
annulus with inner and outer radii 3\,$r_g$ and $R_{\rm d}$, respectively,  with electron column density $N_e$, and magnetic 
field strength $B_d$.  The electron population is assumed to be thermal with temperature $T_e$, where $kT_e\gg m_ec^2$ and $k$, $m_e$ and $c$ are Boltzmann's 
constant, the electron rest mass, and the speed of light, respectively.  For simplicity we neglect the relativistic effects associated with orbital motion and the 
Kerr metric and assume that the disk is viewed face on \citep{gravity18,eht22}.

To constrain the four parameters $R_d,\,N_e,\,T_e, $ and $B_d$, we first ensure that the predicted synchrotron emission matches our VLA measurement of the quiescent 
emission from the disk at 34\, GHz ($1.34\,^{+0.21}_{-0.15}$ Jy, $\S4$), and the existing submm measurements at 233 and 868\,GHz \citep{bower19}, i.e., 2.87, and 
1.86\,Jy respectively.
For a given choice of disk field $B_d$, the other parameters $R_d,\,N_e$ and $\,T_e$ can be chosen to match the four  data points.  Interestingly,  the way in which the synchrotron emission scales with magnetic field leads to a family of disk models parameterized by $B_d$,  
{\it all with an identical spectral energy distribution (SED) $S_\nu$,} as 
plotted in blue in Fig. 11a. 


The model provides a reasonable representation of the emission of Sgr A* above 30\,GHz, but at lower frequencies the emission becomes optically thick 
and $S_\nu\,  \propto\, \nu^{2}$, steeper than the observed $S_\nu\,\propto\,  \nu^{0.3}$ dependence between 1-40 GHz \citep{bower15b}. 
The model under-predicts the flux at lower frequencies because it does not include 
the contribution from the cooler outer part of the accretion flow.  However, this is not a significant shortcoming as seed photons below 30\,GHz make a negligible 
contribution to the X-ray flux and the electrons emitting them are also ineffective in up-scattering NIR seed photons to X-rays.



The  parameters $N_e$, $R_d$ and $kT_e$  depend on powers  of $B_d$
in the following way:
\begin{eqnarray}
    R_d &=& \left(23.94 \, B^{1/2} + 9 \right)^{\!1/2}\, r_g\,,\\
    \label{eqn:Rdisk-scaling}
    N_e &\;=\;& 6.51 \times10^{19} \,B^{-3/2}_\mathrm{d}\,\textrm{cm}^{-2} \,,\textrm{ and}\\
    \label{eqn:Bdisk-scaling}
    kT_e &\;=\;&78.1 \,B^{-1/2}_\mathrm{d}\, \textrm{MeV}\,
    \label{eqn:kTe-scaling}
\end{eqnarray}
where $B_d$ is in units of gauss. 
These dependencies are plotted in Fig. 11b, and the magnitudes of the parameters are in broad agreement with previous attempts to 
match the submm spectrum with one-zone emission models as well as GRMHD simulations of the Sgr A* accretion flow.

Two additional constraints, one observational and one theoretical, limit the range of disk magnetic field $B_d$ to 20--30\,G (see Fig. 11c).   
First, long-term X-ray monitoring of Sgr A* has not detected point-like quiescent emission, finding instead an extended ($1.4''$)  source with a 
spectrum well-fit by thermal brehmstrahlung models, with $L_x\approx 2\times 10^{33}$\,erg\,s$^{-1}$ \citep{baganoff03}. This emission arises on the 
scale of the Bondi radius ($\sim 10^5\,r_g$), and restricts any quasi-steady contribution from the inner flow to $L_X \la 10^{33}$\,erg\,s$^{-1}$.

The disk's synchrotron self-Compton X-ray luminosity, (red curve in Fig. 11c)  strongly rises with decreasing disk 
field strength, so that $B_d$ must be greater than about 20\,G to avoid over-producing quasi-steady X-rays.  
The reason for this strong dependence on $B_d$ is that 
while the density of synchrotron seed photons is fixed by the disk's observed radio-submm SED, the number of electrons in the disk with sufficient energy to 
upscatter them into the X-ray band rises sharply with decreasing $B_d$ because  the number of electrons $\pi R_d^2 N_e$ available to up-scatter them and their temperature 
both increase with decreasing $B_d$.
 
The second constraint is that the ratio of gas pressure to magnetic pressure, the plasma $\beta$, near the midplane of the accretion flow should generally be $\ga 
1$.  We estimate $\beta$ in the disk model assuming that the proton and electron densities are equal, that the proton temperature is ten times that of the 
electrons, and that the thickness of the flow is 0.3 $R_d$.  The resulting $\beta$, plotted as the black line and  in 
Fig. 11b, constrains the disk field in our model to $B_d\la30$\,G.

\subsection{NIR flare source model}

In a similar vein, we model the  NIR flare emission region as a homogeneous sphere, with radius $R_f$, and threaded by a uniform field $B_f$.  The electron energy 
spectrum is a broken power-law chosen  with  a constant rate of injection of electrons with an $E^{-p}$ spectrum that 
are subject to synchrotron losses:  a broken power law, $E^{-p}$ steepening to $E^{-p-1}$ above the energy of the electrons that have a synchrotron loss time of 15 
minutes (halfway through the IR flare). 
The upper and lower cutoff energies are chosen to be 5\,MeV and 1 GeV, respectively, 
and the spectrum is normalized so that the electron and magnetic energy densities are equal.

This model has three parameters: the flare magnetic field strength $B_f$ and radius $R_f$, and the electron energy spectral index $p$.  For given $B_f$, $p$ and $R_f$ are adjusted so 
that the model IR flux matches the two JWST measurements at the peak of the flare, and we again obtain a one-parameter family of solutions, but this time with varying spectral energy 
distributions.  
Earlier  modeling of the JWST light curves for this epoch estimated $B_f\sim\, 88.4$ G \citep{zadeh25}, so by way of illustration, 
Fig. 11a shows the resultant flare spectra for $B_f = 50$ and $100\,$G.  In each case, the spectral turnover occurs where the 
synchrotron optical depth $\sim 1$.

\subsection{Compton up-scattering in channels I, II and III}
    
The disk and flare source models, as shown schematically in Figure 10 and quantitatively in Figure 11a,b, specify the electron populations and the photons available 
to participate in Compton up-scattering.  
A geometric relationship between the two  is also required so that the density of disk-emitted photons in the flare 
source and the density of flare produced photons in the disk can both be determined. Again, for simplicity, we place the flare source on the disk axis at height 
5\,$r_g$.


In calculating the X-ray emissivity generated by Compton up-scattering, we make the simplifying approximation that an incident seed IR photon of energy $h\nus$ is up-scattered by an 
electron with Lorentz factor $\gamma$ to energy $\frac{4}{3}\gamma^2h\nus$.  We also account for the synchrotron absorption of submm seed photons within the flare source.

\begin{figure}
    \centering
    \includegraphics[width=12.5cm]{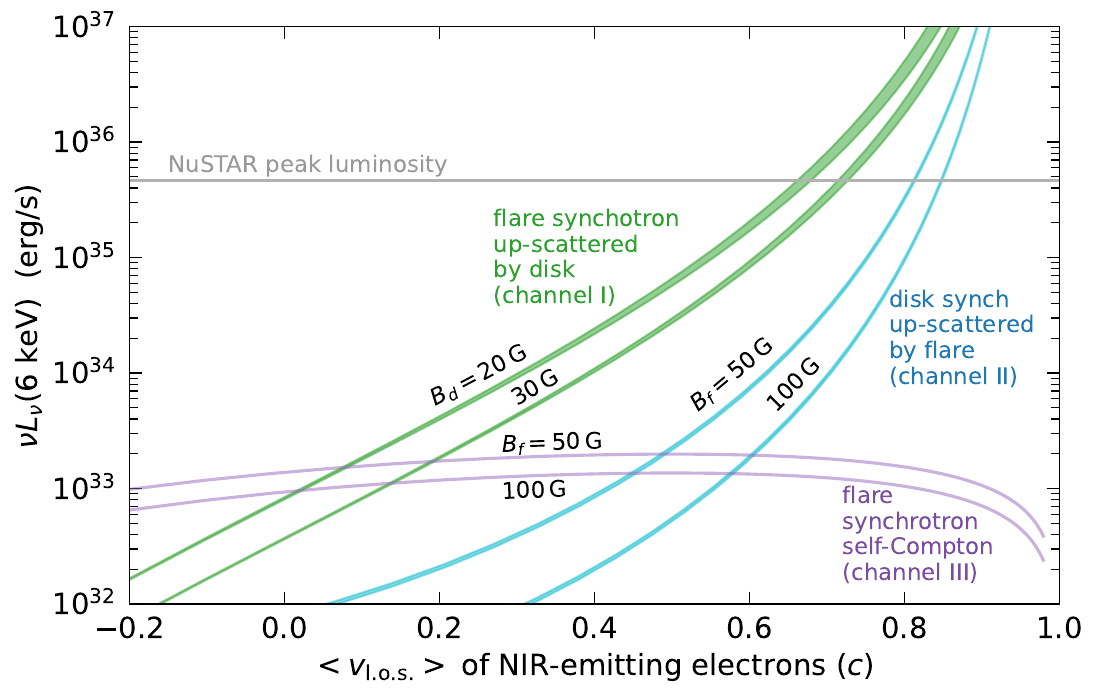}
\caption{
Predicted X-ray flare luminosities arising from Compton up-scattering of disk and flare synchrotron photons as a function of the line-of-sight bulk velocity of the IR emitting 
flare source:  IR 
flare photons by thermal electrons in the disk (green), submm photons from the disk by the non-thermal electrons in the IR flare source (cyan), and synchrotron self-Compton in the IR 
flare source (purple).  The gray line indicates the peak X-ray luminosity detected by NuSTAR: this can only be achieved if the  IR-emitting electrons 
are moving away from the observer and toward the disk with bulk  speed $v\sim0.7\,c$. 
The thickness of the Channel I and II   curves show the flare source size  dependence 
 (See text for further explanation).
\label{fig:F12_Lx-vs-beaming}
}
\end{figure}

Our results are summarized in Fig.~\ref{fig:F12_Lx-vs-beaming}.  If the flare source is motionless, corresponding to $v=0$ in Fig.~\ref{fig:F12_Lx-vs-beaming}, none 
of the three channels are able to contribute more than about $10^{33}$\,erg\,s$^{-1}$ to $L_X$, two and a half orders of magnitude below the NuSTAR peak flux.

 \begin{figure}
    \centering
    \includegraphics[width=12.5cm]{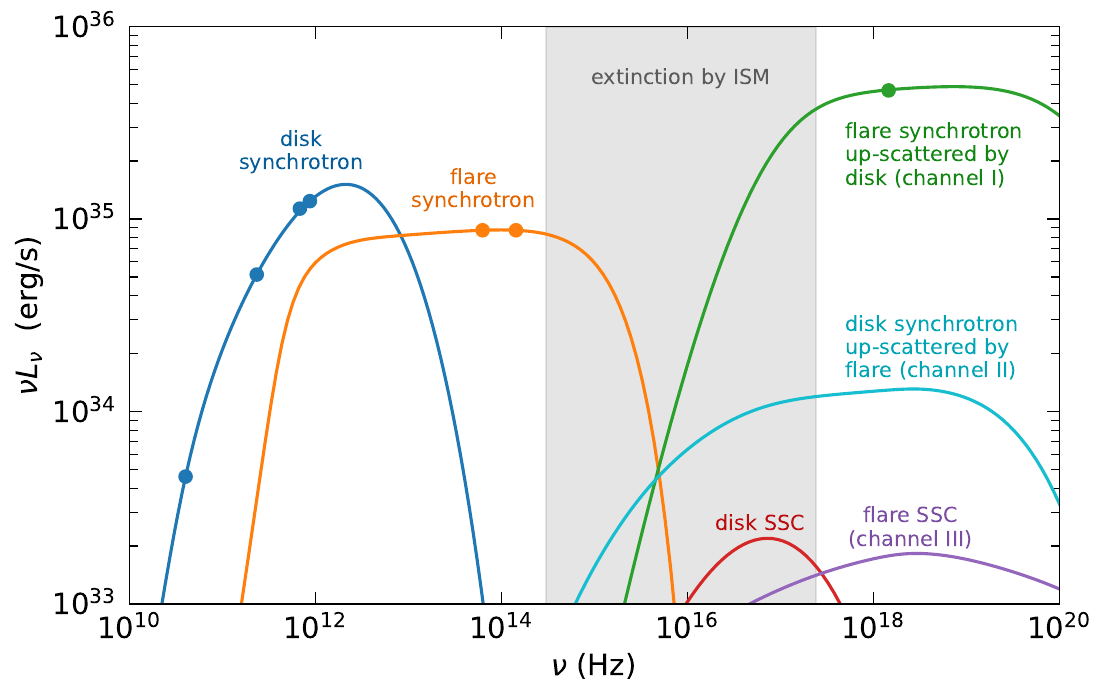}
\caption{
Model broadband spectrum of Sgr A* 
during the peak of the JWST/NuSTAR flare.  The blue 
curve shows the quasi-steady thermal synchrotron emission from the accretion flow $(``disk")$ with points indicating the radio sub-mm measurements. 
The synchrotron emission from the 
NIR flare is plotted in orange with the JWST points indicated.  Green, cyan and purple shows the corresponding X-ray flare emission arising from 
Compton up-scattering via channels I, 
II and II respectively at v$_{los}$=0.7c (see text); the green point indicates the NuSTAR X-ray flux at 6\,keV.  The gray shaded region indicates the range of frequencies (1\,$\mu$m to 1\,keV 
equivalent) that are not accessible to observation  because of interstellar extinction.
\label{fig:F13_broadband-spectrum}
}
\end{figure}


We therefore allow the electron population in the emission region  to have a bulk  velocity  towards or away from the observer and consider how beaming effects modify the predicted flux in all three channels. We 
focus on the most favorable case in which the bulk motion is directed away 
 from the observer and towards the disk.  In this picture, relativistic beaming reduces the 
observed IR flux, so that the intrinsic luminosity of the emitting plasma  at NIR is higher than would be inferred if  it were static.  Further, the IR emission is 
beamed towards the disk, and the disk emission appears beamed towards the flare, thus the disk emission is similarly enhanced. In 
both channels I and II,  the net effect is to significantly increase the number density and energy of the seed photons available for up-scattering to X-rays. 



The predicted X-ray fluxes from each of the three channels as a function of the line-of-sight bulk  velocity of the emitting region in the flare source are shown in 
Fig.~\ref{fig:F12_Lx-vs-beaming}. 
The channel III (flare SSC, purple curve) luminosity is roughly constant
since there is no relative motion between the source and scatterers. 
 Meanwhile, the predicted luminosity from ICS channels I (green) and II (cyan) both increase strongly with velocity, as expected from the discussion in 
the previous paragraph.  Channel I dominates II because (a) the disk is optically thin to the incident IR seed photons from the flare source, whereas many of the 
submm seed photons incident from the disk suffer synchrotron absorption before being up-scattered, and (b) the disk subtends a larger solid angle as seen from 
 the flare than 
the flare source does as seen from  the disk.

The channel I luminosity decreases with increasing disk field $B_d$ because the column of electrons in the disk needed to maintain the disk's observed synchrotron 
luminosity is lower, (see Fig. 11b), reducing the rate of up-scattering in the disk.  The thickness of each channel I curve reflects 
the weak dependence of the IR seed photon field on the change in flare source size.  Similarly, the channel II X-ray luminosity weakly increases with increasing 
$B_d$ because of  
the decreased  solid angle subtended by the disk (see Eq. 5) as viewed by the flare source. 

In Fig. 13 we show the contributions of different emission components to the broadband spectrum of Sgr A* at the peak of the NIR/X-ray flare, assuming $B_d = 30$\,G, $B_f=100$\,G and 
that the NIR-emitting electrons have bulk motion of $\sim0.7c$ towards the disk.  The disk contributes two quasi-steady components arising from thermal synchrotron emission (blue) and 
synchrotron self-Compton emission (red).  The SSC emission is less than $10^{33}$ erg s$^{-1}$ in the X-ray as required by the Chandra long-term monitoring constrain.  The NIR flare component 
(orange) engenders the three X-ray flaring components corresponding to channels I—III.  As expected, the X-ray luminosity is dominated by Compton up-scattering of NIR flare photons 
by the accretion flow (channel I).

\section{A Physical picture of X-ray, IR and radio variability}

We suggest a dynamic picture that unifies flux variability of Sgr A* at X-ray, IR and radio/submm wavelengths. In this picture, the acceleration of particles 
producing NIR emission is due to magnetic reconnection at the X-point of a current sheet.  As for X-rays, we conclude that the NuSTAR flare could arise from ICS of IR 
flare photons from the disk (i.e.~channel I)  (see Figure \ref{fig:F12_Lx-vs-beaming}), if the IR-emitting electrons have bulk motion  $\sim 0.7\,c$ away from the 
observer and towards the disk. However, as $v\sim\, 0.7\,c$  is equivalent to displacement by $1\, r_g$ every 30 seconds and the IR flare duration is 40 minutes. These are 
unlikely to be true physical displacement of an IR-emitting source. We suggest instead that the IR flare source is the ``downward'' or ``diskward'' component of 
particles accelerated during a reconnection event occurs  below a flux rope that has been ejected from the disk \citep{lin24}. 

A  diagram of the  physical picture that  we envision  is displayed in  Figure 14. 
Reconnection occurs  in a current sheet extending from a magnetic flux rope,  which is ejected towards the observer,  
back to the accretion flow of Sgr A* (orange), analogous to the reconnection 
topology of solar flares  
\citep{chen20,drake25}.  
Plasma flows into the reconnection point (pink arrows) where it is heated, 
seeded with accelerated electrons, then ejected parallel to the current sheet 
(red arrows).  The energetic electrons directed towards the accretion flow create a region of NIR synchrotron emission, whereas 
those directed towards the observer experience lower magnetic field 
strengths and so emit appreciable synchrotron emission at longer wavelengths including sub-mm (blue shading) and radio frequencies (dark orange).
We emphasize that despite the bulk motion of the emitting plasma, 
there is {\it no bulk motion}  of the accelerated particles in the reconnecting region. As  reconnection proceeds, plasma containing  a mixture of newly-accelerated electrons 
is continually 
ejected  from the X-point at the Alfven speed, $\sim$0.7c.   
Half of the 
accelerated electrons are pushed downwards towards the disk at  $v\,\sim 0.7\,c$, radiating away their energy on the 
synchrotron loss time scale, $\sim 1.4\,{\rm min}$ for $B_f\sim 100\,$G  yielding a  source size   of $\sim 3\,r_g$.


The remaining half of the electrons are accelerated away from  the disk where the magnetic field is expected to be lower 
and  is injected into the region surrounding the magnetic flux rope. Radio emission in this picture is delayed with respect to the NIR flare emission because of 
higher opacity of  radio emission at low frequencies. 
Numerical simulations of accretion flows show
that turbulent motion and differential rotation of the accretion flow  produce flares and flux ropes 
\citep{markoff01,yuan03,igumenshchev08,yuan09,dodds-eden10,dibi14,dexter20,ripperda20,ressler20a,ball21,porth21,cemeljic22,aimar23,lin24}. 

 
Flux rope ejection  should occur equally on both sides of the accretion disk.  In the case where reconnection  occurs on the far-side,  the disk  
 $``sees"$ exactly what the  observer sees, and  there is no net gain  from beaming compared to the static case. 
In our picture we expect half the reconnection events to be on the far  side of the disk, with little  X-ray emission.  




\begin{figure}
    \centering
    \includegraphics[width=12.5cm,angle=0]{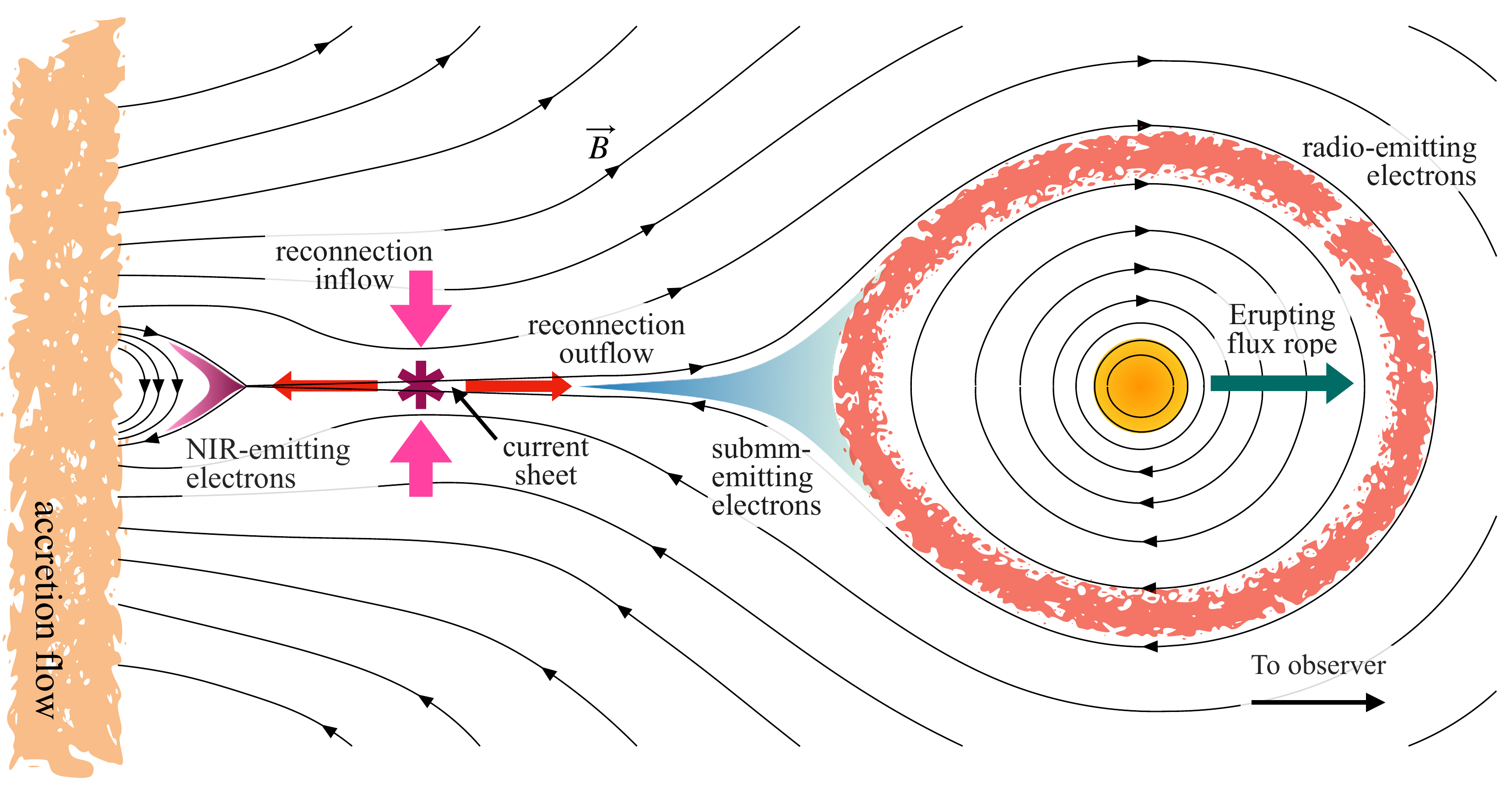}
\caption{
A Schematic diagram of how X-ray flare emission from Sgr A* is generated by ICS of NIR photons. 
This picture is very similar to production of 
solar flare emission and coronal mass ejection from the Sun \citep{chen20,drake25}, also see \citep{lin24}.  
Reconnection occurring in a current sheet extending from a flux rope ejected towards the observer back to the accretion flow of Sgr A* (orange)
\citep{chen20}.  Heated  plasma flows into the reconnection point (pink arrows), followed  by 
ejection  parallel to the current sheet 
(red arrows).  
The energetic electrons directed toward the accretion flow  produce  NIR synchrotron emission,
whereas those away from the disk 
 emit appreciable synchrotron emission at sub-mm (blue shading) and radio frequencies (dark orange) (see text).
} 
\label{fig:F14_broadband-spectrum}
\end{figure}

%
%
%
%

\vfill\eject
\section{Summary}

We presented simultaneous IR  and X-ray flare emission and  delayed variable radio emission  from Sgr A* on 
April 5, 2024.  We modeled X-ray flare emission and concluded that only the ICS mechanism can explain the bright X-ray flare 
and that there is also a physical connection between NIR and radio/submm variable emission whenever there is an outburst from the accretion disk of Sgr A*, at least 
for this and the other most luminous detected X-ray flares \citep{porquet08,nowak12,zhang17,haggard19}.

Our observations show the spectral evolution of NIR, and  radio variable emission as well as the spectrum of  X-ray flare emission, 
thus constraining emission models.  
We proposed a new picture that unifies flare emission at IR, X-ray and submm/radio wavelengths. 
Flaring events  driven by 
reconnection are consistent with  simulations 
\citep{markoff01,yuan03,igumenshchev08,yuan09,dodds-eden10,dibi14,dexter20,ripperda20,ressler20a,ball21,porth21,cemeljic22,aimar23,lin24}.
In this scenario 
particles are accelerated by reconnection and magnetic flux ropes  are expelled from the disk in a topology that resembles a solar flare \citep{chen20,drake25}, analogously 
producing oppositely-directed streams of energetic particles moving towards and away from the accretion disk \citep{lin24}. 
 NIR synchrotron radiation from the relativistic electrons 
moving toward the disk is beamed towards and upscattered by thermal electrons in the disk, producing the observed strong X-ray flare.  Meanwhile, electrons 
propagating away from the reconnection site are injected into an  escaping magnetic flux rope and because of the reduced field strength emit in the FIR and at 
later times in the submm  and radio as the rope expands.





\section*{Acknowledgments}
We thank S. Ressler and V. Tatischeff for insightful discussions.  Work by R.G.A. was supported by NASA under award 
number 80GSFC24M0006. Work by FYZ was supported by the grant AST-2305857 from the NSF and NASA under awards JWST-2235 and JWST-3559.
JMM is supported by an NSF Astronomy and Astrophysics Postdoctoral Fellowship under award AST-2401752.
JWST observations are associated with NASA's JWST-2235 and JWST-3559 programs. Some of the data 
presented in this Letter were obtained from the Mikulski Archive for Space Telescopes (MAST) at the Space Telescope Science Institute. All the JWST data used in this Letter can be 
found in MAST: doi:10.17909/0z7e-gf46. This work is based on observations made with the NASA/ ESA/CSA James Webb Space Telescope. The data were obtained from the Mikulski Archive for 
Space Telescopes at the Space Telescope Science Institute, which is operated by the Association of Universities for Research in Astronomy, Inc., under NASA contract NAS 5-03127 for 
JWST. The National Radio Astronomy Observatory is facility of the U.S. National Science Foundation operated under cooperative agreement by Associated Universities, Inc. COH is 
supported by NSERC Discovery Grant RGPIN-2023-04264, and Alberta Innovates Advance Program 242506334.

\end{document}